\documentclass[journal]{IEEEtran}
\usepackage[colorlinks,linkcolor=magenta,anchorcolor=magenta,citecolor=magenta,bookmarks=true]{hyperref}  
\usepackage[T1]{fontenc}
\ifCLASSINFOpdf
\usepackage[pdftex]{graphicx} 
\DeclareGraphicsExtensions{.eps,.pdf,.jpeg,.png}
\else
\usepackage[dvips]{graphicx}
\fi
\usepackage[compress,nospace]{cite}
\usepackage[cmex10]{amsmath}
\usepackage{epstopdf,amsthm,stfloats,siunitx,amssymb,wasysym,algorithm,algorithmic,array,url, color,subfigure} 
\usepackage{footnote}
\makesavenoteenv{tabular}
\usepackage{soul}
\newcommand{\tabincell}[2]{
	\begin{tabular}{@{}#1@{}}#2\end{tabular}
}
\begin{document}

%% Title and Authors
\title{SDOA-Net: An Efficient Deep Learning-Based DOA Estimation Network for Imperfect Array}

\author{Peng~Chen,~\IEEEmembership{Senior~Member,~IEEE}, Zhimin~Chen,~\IEEEmembership{Member,~IEEE}, Liang~Liu,~\IEEEmembership{Senior~Member,~IEEE}, Yun~Chen,~\IEEEmembership{Senior~Member,~IEEE}, 
Xianbin~Wang,~\IEEEmembership{Fellow,~IEEE}

\thanks{This work was supported in part by the Natural Science Foundation for Excellent Young Scholars of Jiangsu Province under Grant BK20220128, the Natural Science Foundation of Shanghai under Grant 22ZR1425200,  the Open Fund of State Key Laboratory of Integrated Chips and Systems under Grant SKLICS-K202305, the National Key R\&D Program of China under Grant 2019YFE0120700, the Open Fund of National Key Laboratory of Wireless Communications Foundation under Grant IFN20230105, the Open Fund of National Key Laboratory on Electromagnetic Environmental Effects and Electro-optical Engineering under Grant JCKYS2023LD6, the Open Fund of ISN State Key Lab under Grant ISN24-04, and the National Natural Science Foundation of China under Grant 61801112. \textit{(Corresponding author: Zhimin Chen)}} 
 
\thanks{Peng~Chen is with the State Key Laboratory of Millimeter Waves, Southeast University, Nanjing 210096, China, and also with State Key Laboratory of Integrated Chips and Systems, Fudan University, Shanghai 201203, China (e-mail: chenpengseu@seu.edu.cn).}
\thanks{Zhimin~Chen is with the School of Electronic and Information, Shanghai Dianji University, Shanghai 201306, China, and also with the Department of Electronic and Information Engineering, The Hong Kong Polytechnic University, Hong Kong (e-mail: chenzm@sdju.edu.cn).}
\thanks{Liang~Liu is with the Department of Electronic and Information Engineering, The Hong Kong Polytechnic University, Hong Kong (e-mail: liang-eie.liu@polyu.edu.hk).}
\thanks{Yun~Chen is with the State Key Laboratory of Integrated Chips and Systems and the Microelectronics School, Fudan University, Shanghai 201203, China (e-mail: chenyun@fudan.edu.cn).}
\thanks{Xianbin~Wang is with the Department of Electrical and Computer Engineering, Western University, London, ON N6A 5B9, Canada (e-mail: xianbin.wang@uwo.ca).} 
}

\markboth{IEEE Transactions on Instrumentation and Measurement}%
{Shell \MakeLowercase{\textit{et al.}}: Bare Demo of IEEEtran.cls for Journals}

\maketitle

\begin{abstract}
The estimation of direction of arrival (DOA) is a crucial issue in conventional radar, wireless communication, and integrated sensing and communication (ISAC) systems. However, low-cost systems often suffer from imperfect factors, such as antenna position perturbations, mutual coupling effect, inconsistent gains/phases, and non-linear amplifier effect, which can significantly degrade the performance of DOA estimation. This paper proposes a DOA estimation method named super-resolution DOA network (SDOA-Net) based on deep learning (DL) to characterize the realistic array more accurately. Unlike existing DL-based DOA methods, SDOA-Net uses sampled received signals instead of covariance matrices as input to extract data features. Furthermore, SDOA-Net produces a vector that is independent of the DOA of the targets but can be used to estimate their spatial spectrum. Consequently, the same training network can be applied to any number of targets, reducing the complexity of implementation. The proposed SDOA-Net with a low-dimension network structure also converges faster than existing DL-based methods. The simulation results demonstrate that SDOA-Net outperforms existing DOA estimation methods for imperfect arrays. The SDOA-Net code is available online at \url{https://github.com/chenpengseu/SDOA-Net.git}.
\end{abstract}


\section{Introduction} \label{sec1}
\IEEEPARstart{D}{irection} of arrival (DOA) estimation is a fundamental problem in wireless communications, radar-based applications, and future integrated sensing and communication (ISAC) systems~\cite{10005294,zhang_overview_2021,9845366,xu_rate-splitting_2021}, and has been studied for decades. Typically, the DOA estimation is based on an ideal antenna array model, without considering any imperfect effect, including the mutual coupling effect, inconsistent gains/phases, non-linear effect, etc. In this ideal scenario, DOA can be estimated by traditional methods such as the monopulse angle estimation method~\cite{zhu_combined_2018} and fast Fourier transformation (FFT)-based methods. 

In addition, there have been proposals for super-resolution estimation methods. Subspace-based methods, such as multiple signal classification (MUSIC)~\cite{lin_fsf_2006,yan_low-complexity_2013,zhang_direction_2010} and estimation of signal parameters via rotational invariance techniques (ESPRIT)~\cite{kim_joint_2015,lin_time-frequency_2016,xiaofei_zhang_multi-invariance_2009,fang-ming_han_esprit-like_2005}, have been suggested.  An optimization problem is formulated in~\cite{10007797} to estimate the DOA considering the eigenvalues ranking problem. In addition, sparse reconstruction-based methods have been introduced that take advantage of the sparsity of signals in the spatial domain. For example, compressed sensing (CS)-based methods have been proposed for DOA estimation, including sparse Bayesian learning-based methods~\cite{10041189,wan_deep_2021,p_chen_off-grid_2019,10172104,Dai_So_2021,mao_marginal_2021,wan_robust_2022,wang_alternative_2018} and mixed $\ell_{2,0}$ norm-based methods~\cite{5466152}.  
 
However, the above works did not consider the effect of imperfect antenna arrays. As a result, the performance of these algorithms is significantly affected in practical DOA estimation systems. In the literature, some work has started to investigate DOA estimation schemes under imperfect antenna arrays. For example, for an array with mutual coupling, gain or phase errors, and sensor location errors, a method for estimating DOA and model errors is proposed in~\cite{lu_direction--arrival_2018}. A fourth-order parallel factor decomposition model using imperfect waveforms is given in~\cite{ruan_parafac_2019} to estimate the DOA. Then, ref.~\cite{liu_2-d_2021} proposes a two-dimensional (2D) DOA estimation method for an imperfect L-shaped array using active calibration. However, each of the above works only considered a subset of the imperfect array effects because optimization over the complicated array model with all imperfect effects considered is challenging. This motivates us to use the deep learning (DL) technique for DOA estimation with all imperfect array effects taken into account because of its efficiency for training over difficult networks.

In the literature, several works have been done for DL-based DOA estimation~\cite{9173575,8400482,7497454,9178434}, and they have the advantages of low computational complexity and high accuracy. There are some types of DL-based methods:
\begin{enumerate}
	\item The input data is the raw sampled data from the array;
	\item The input data is the covariance matrix of the received signal;
	\item The outputs are the directly estimated DOAs;
	\item The output is the spatial spectrum and the DOAs are estimated from the spectrum.
\end{enumerate}
Various DL-based methods have been proposed for DOA estimation in the literature. In~\cite{yuan_unsupervised_2021,wu_deep_2019}, the input is the covariance matrix, the output is the spectrum, and a sparse loss function is used to train the network. Ref.~\cite{papageorgiou_deep_2021} uses the estimated covariance matrix as input and discretizes the spatial domain into grids to estimate the DOA. A synthetic dataset is shown in~\cite{akter_rfdoa-net_2021}, and a CNN-based method is proposed for the estimation of DOA in the presence of additive noise, propagation attenuation, and delay. For coherent signals, an angle separation learning method is proposed in~\cite{wu_deep_2019}, and the covariance matrix is formulated as input features of the DNN. In~\cite{8854868}, a deep convolution network (DCN) is given for DOA estimation with the covariance matrix as the undersampled linear measurements of the spatial spectrum, where the signal sparsity in the spatial domain is also exploited to improve estimation performance. A MUSIC-based DOA estimation method is proposed in~\cite{9328861} using small antenna arrays, where DL is formulated to reconstruct the signals of a virtual large antenna array. Ref.~\cite{8400482} gives an offline and online DNN method for the estimation of DOA in the massive multiple input multiple output (MIMO) system, where DOA is the network output and can be estimated directly from the received signal. For the estimation of DOA with a low signal-to-noise ratio (SNR), a convolutional neural network (CNN) is proposed in~\cite{9457195}, where the covariance matrix is the input of the network and shows increased robustness in the presence of noise. Moreover, a multiple deep CNN is designed in~\cite{9034077}, where each CNN learns the MUSIC spectrum of the received signal, so a non-linear relationship between the received sensor data and the angular spectrum is formulated in the network. For the imperfect array, ref.~\cite{8485631} introduces a DNN framework to estimate the DOA using a multi-task autoencoder and a series of parallel multilayer classifiers. 
 
We find that the DL-based DOA estimation methods mainly use CNN as a typical network structure~\cite{chakrabarty_multi-speaker_2019}, and the input is the statistic results such as the covariance matrix. Since the information in the statistical data limits the estimation performance, the performance cannot be better using the raw sampled data. Furthermore, the network output is the estimated DOAs, and the spatial spectrum cannot be obtained. Therefore,  the network structure should be adjusted with different target numbers and is not suitable for practical applications. There are some limitations to existing DL-based methods:
\begin{itemize}
	\item Since the classic DOA estimation algorithms, such as MUSIC, is just based on the covariance matrices of the received signals, most existing ML-based schemes use these covariance matrices as the input data to train the network. However, the covariance matrices are not sufficient for the optimal estimator design, in general. As a result, the input data used in these works do not preserve all useful information.  
	\item Furthermore, the output of existing ML-based DOA estimation schemes is usually the spatial spectrum of the targets. In this case, the training network depends on the number of targets; that is, different networks should be trained given a different number of targets. This is of high complexity in practice. 
	\item Furthermore, when the spatial spectrum is used, we must discretize the DOAs into grids, and the possible DOA must be on the discretized grids exactly. More girds as the output must be used for high accuracy, and the network will become more complex and difficult to train.
\end{itemize}

In this paper, we propose a DNN network based on CNN, i.e., super-resolution DOA network (SDOA-Net), to overcome the above-mentioned difficulties in the DOA estimation. The proposed SDOA-Net is used for the performance evaluation of imperfect arrays under realistic conditions. Compared with existing methods, the proposed SDOA-Net can achieve better estimation performance with lower complexity. The contributions of this paper are given as follows:
\begin{enumerate}
	\item \textbf{A system model with imperfect array effects for the DOA estimation is formulated.} The imperfect effect includes the position perturbation, the inconsistent gains, the inconsistent phases, the mutual coupling effect, and the nonlinear effect, etc. 
	As a result, our results are directly applicable to a practical system.
	\item \textbf{A DL architecture is proposed based on the imperfect array.} Unlike existing methods, the input of SDOA-Net is the raw sampled signals and the output is a vector, which can be easily used to estimate the spatial spectrum. Convolution layers are then used to get the signals' features and avoid the complexity of high-dimension signals. The SDOA-Net output is a vector for the spectrum estimation and can avoid the problem of discretizing the spatial domain. Compared to the existing CNN-based method, the proposed SDOA-Net can be easily trained and perform the better estimation. 
	\item \textbf{A spatial spectrum-based loss function to train the SDOA-Net is proposed,} where Gaussian functions are used to approximate the spatial spectrum. Inspired by the atomic norm minimization (ANM)-based DOA estimation method, the output of SDOA-Net is used to formulate the spatial spectrum. Differently from existing networks for the DOA estimation, we use a special spectrum-based loss function to measure the error between the reference spectrum and the estimated one and to train the network.
\end{enumerate}

The remainder of this paper is organized as follows. The system mode of practical DOA estimation is formulated in Section~\ref{sec2}. The review of the super-resolution DOA estimation method is given in Section~\ref{sec3}. Then, the proposed SDOA-Net for DOA estimation is shown in Section~\ref{sec4}. The simulation results are carried out in Section~\ref{sec5}, and finally Section~\ref{sec6} concludes the paper. 

\textit{Notations:}  Matrices and column vectors are denoted by upper- and lower-case boldface letters, respectively. The matrix transpose and the Hermitian transpose are represented by $(\cdot)^\text{T}$ and $(\cdot)^\text{H}$, respectively. The real and imaginary parts of a complex value are denoted by $\mathcal{R}{\cdot}$ and $\mathcal{I}{\cdot}$, respectively. The trace of a matrix is denoted by $\text{Tr}{\cdot}$. The $\ell_2$ norm is represented by $|\cdot|_2$.

\section{The System Model for Practical DOA Estimation}\label{sec2}

\begin{figure}
	\centering
	\includegraphics[width=3in]{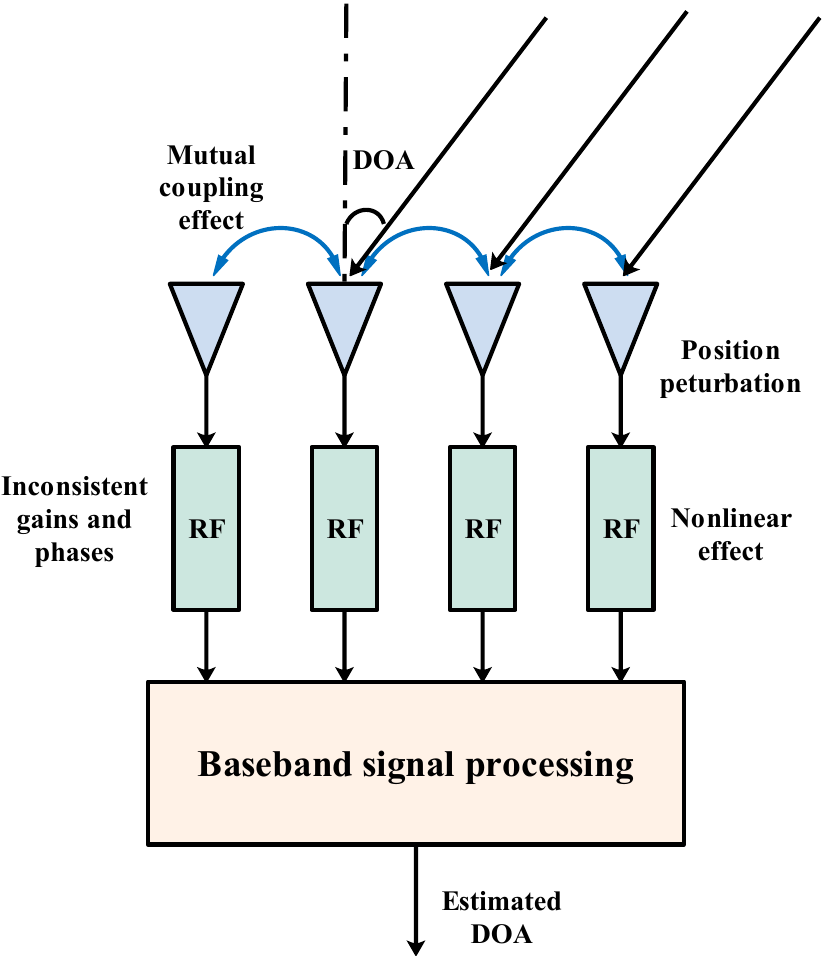}
	\caption{The system model for the DOA estimation in a practical array.}
	\label{sys}
\end{figure}

In this paper, we consider the DOA estimation problem in a practical system and propose a DL-based estimation framework. As shown in Fig.~\ref{sys}, we consider $K$ far-field signals, and the $k$-th ($k=0,1,\dots,K-1$) signal is expressed as $s_{k}(t)\in\mathbb{C}$ with the DOA being $\theta_k\in\left(-\frac{\rm{\pi}}{2},\frac{\rm{\pi}}{2}\right]$. A linear array system with $N$  antennas is used to receive the signals and estimate the DOAs, where the wavelength is denoted as $\lambda$. Taking into account the additive noise $w_n(t)\in\mathbb{C}$, the signal received at the $n$-th ($n=0,1,\dots, N-1$) antenna can be expressed as
\begin{align}\label{eq1}
	r_n(t) = g\left(x_n(t)+\sum_{n'\neq n}B_{n,n'}x_{n'}(t)\right)+w_n(t).
\end{align}
Then, we have
\begin{align}
	x_n(t) = \sum^{K-1}_{k=0}s_k(t)A_n\text{e}^{j\phi_n}\text{e}^{j2\rm{\pi}\frac{d_n}{\lambda}\sin\theta_k},
\end{align}
where taking the $0$-th antenna as the reference one, i.e., $d_0=0$, the position of the $n$-th antenna is $d_n$, and for a uniform linear array (ULA), the position of the antenna is $d_n=n\lambda/2$. In the received signal (\ref{eq1}), the following imperfect problems are considered:
\begin{enumerate}
	\item \emph{The mutual coupling effect:} The antennas cannot be ideally isolated and introduce the mutual coupling effect among the received signals. The mutual coupling coefficient between the $n$-th and $n'$-th ($n\neq n'$) antenna is $B_{n,n'}\in\mathbb{C}$ with $|B_{n,n'}|<1$ in (\ref{eq1});
	\item \emph{The position perturbations:} The antenna positions cannot be exactly at the desired positions, and will cause the phase errors of the received signals among antennas in the steering vector;
	\item \emph{The inconsistent gains: } The radio frequency (RF) channels usually cannot have exactly the same amplifiers, and will cause amplitude differences among the received signals. The channel gain of the $n$-th antenna is denoted as $A_n>0$;
	\item \emph{The inconsistent phases: } The difference among the RF channels will also cause the delay and phases errors of the received signals, and The channel phase of the $n$-th antenna is denoted as $\phi_n$;
	\item \emph{The nonlinear effect: } The nonlinear effect among RF channels and analog-to-digital converter (ADC) will introduce the nonlinear effect and degrade the DOA estimation performance.  We use a non-linear function $g(\cdot)$ to represent the nonlinear operation in the receiving channels.
\end{enumerate}
 
Hence, collect the received signals into a vector 
\begin{align}
\boldsymbol{r}\triangleq \begin{bmatrix}
	r_0(t), r_1(t),\dots, r_{N-1}(t)
\end{bmatrix}^{\text{T}}.	
\end{align}
The DOA estimation problem can be formulated as a  parameter estimation problem with the received signal $\boldsymbol{r}$. Most existing works consider the methods in the scenario with the perfect array, where we have the linear function $g(\cdot)$, the mutual coupling coefficient $B_{n,n'}$ is $0$, the channel gains are the same ($A_n=1$ and $\phi_0=0$), and the position $d_n$ of the antenna is known. 

However, when an imperfect array is considered, the imperfect elements include the mutual coupling effect, the nonlinear effect, the inconsistent phases, the inconsistent gains, and the position perturbations, etc. In the practical systems, most existing super-resolution methods cannot outperform the traditional methods, where the super-resolution methods must have perfect systems and high SNR. In this paper, we will focus on a robust super-resolution method for DOA estimation with imperfect system effects. 

\section{The Review of Super-Resolution DOA Estimation Methods}\label{sec3}
\subsection{The Atomic Norm-Based Estimation Methods}
In recent years, atomic norm-based methods have been proposed for line-spectra estimation and achieved better performance by exploiting the sparsity of the spectrum in the frequency domain. Additionally, the DOA estimation problem can be easily described as a line-spectral estimation problem, so atomic norm-based methods have been proposed for the DOA estimation.

Usually, in the atomic norm-based methods. the ideal ULA is assumed, and the received signal based on (\ref{eq1}) in the $n$-th array can be expressed as
\begin{align}
	r_n=\sum^{K-1}_{k=0}s_k\text{e}^{j2\rm{\pi}\frac{n d}{\lambda}\sin\theta_k} + w_n(t),
\end{align}
where the distance between adjacent antennas is $d=\frac{\lambda}{2}$. Then, with the definition of a steering vector
\begin{align}
	\boldsymbol{a}(\theta)\triangleq \begin{bmatrix}
	1, \text{e}^{j\frac{2\rm{\pi} d}{\lambda}\sin\theta},\dots, 1, \text{e}^{j\frac{2\rm{\pi} (N-1)d}{\lambda}\sin\theta}
\end{bmatrix}^{\text{T}},
\end{align}
collect all the received signals into a vector, and we have
\begin{align}
	\boldsymbol{r}&\triangleq \begin{bmatrix}
		r_0,r_1,\dots,r_{N-1}
	\end{bmatrix}^{\text{T}}\notag \\
	& = \boldsymbol{As}+\boldsymbol{w},
\end{align} 
where we define the steering matrix as
\begin{align}
	\boldsymbol{A}&\triangleq \begin{bmatrix}
		\boldsymbol{a}(\theta_0),\boldsymbol{a}(\theta_1),\dots, \boldsymbol{a}(\theta_{K-1})
	\end{bmatrix},
\end{align}
the signal vector is defined as
\begin{align}
	\boldsymbol{s}&\triangleq \begin{bmatrix}
		s_0,s_1,\dots,s_{K-1}
	\end{bmatrix}^{\text{T}},
\end{align}
and the noise vector is
\begin{align}
	\boldsymbol{w}&\triangleq \begin{bmatrix}
		w_0,w_1,\dots,w_{N-1}
	\end{bmatrix}^{\text{T}}.
\end{align}

In the ANM-based DOA estimation method, an atomic norm is defined as 
\begin{align}
	\|\boldsymbol{x}\|_{\mathcal{A}}\triangleq \inf \Big\{ & \sum_n \alpha_{n'}:\boldsymbol{x}=\sum \alpha_{n'} \text{e}^{j\phi_{n'}} \boldsymbol{a}(\theta_{n'}), \notag\\
	& \phi_{n'}\in[0,2\rm{\pi}), \alpha_{n'}\geq 0
	\Big\},  
\end{align}
which describes a sparse representation  of $\boldsymbol{x}$ with the sparse coefficients being $\alpha_{n'}$ ($n'=0,1,\dots, N'-1$). Then, with the received signal $\boldsymbol{r}$, we denoise the signal with a sparse reconstruction signal $\boldsymbol{x}$, which can be expressed as an ANM expression 
\begin{align}
	\min_{\boldsymbol{x}} \frac{1}{2}\|\boldsymbol{r}-\boldsymbol{x}\|^2_2+\beta \|\boldsymbol{x}\|_{\mathcal{A}},
\end{align}
where the parameter $\beta$ is used to control the trade-off between the sparsity and the reconstruction accuracy. This ANM problem can be solved by introducing a semi-definite programming (SDP) method, which is
\begin{align}\label{eq11}
	\min_{\boldsymbol{B},\boldsymbol{h}}\quad & \|\boldsymbol{r}-\boldsymbol{h}\|^2_2\notag\\
	\text{s.t}\quad & \begin{bmatrix}
		\boldsymbol{B}& \boldsymbol{h}\\
		\boldsymbol{h}^{\text{H}}& 1
	\end{bmatrix}\succeq 0\\
	& \boldsymbol{B}\text{ is Hermitian matrix}\notag\\
	& \text{Tr}\{\boldsymbol{B}\} = \beta^2\notag\\
	& \sum_n \boldsymbol{B}_{n,n+n'} =0,\text{ for } n'\neq 0 \notag \\ &\qquad \qquad\qquad \quad \text{ and }n'=1-N,\dots,N-1.\notag
\end{align}
By solving the SDP problem (\ref{eq11}), the sparse reconstruction signal $\boldsymbol{h}$ can be obtained, and the DOA of the received signal can be estimated by finding the peak values of the following polynomial 
\begin{align}
	f(\theta) = |\boldsymbol{a}^{\text{H}}(\theta)\boldsymbol{h}|^2.
\end{align}

The ANM-based DOA estimation method is for the ideal array with perfect assumptions, but for the practical array, the ANM-based method must be extended. In~\cite{p_chen_new_2020,wang_gridless_2019,govinda_raj_single_2019-1,gong_doa_2022}, the atomic norm-based methods are extended for the practical array. We can find that the much complex optimization problems are formulated, and a vector like $\boldsymbol{h}$ denoted as $\boldsymbol{h}'$ can be obtained. Then, the DOAs are estimated by the peak values of the following polynomial 
\begin{align}
	f'(\theta) = |\boldsymbol{a}^{\text{H}}(\theta)\boldsymbol{h}'|^2.
\end{align} 

\begin{figure*}
	\centering
	\includegraphics[width=7.2in]{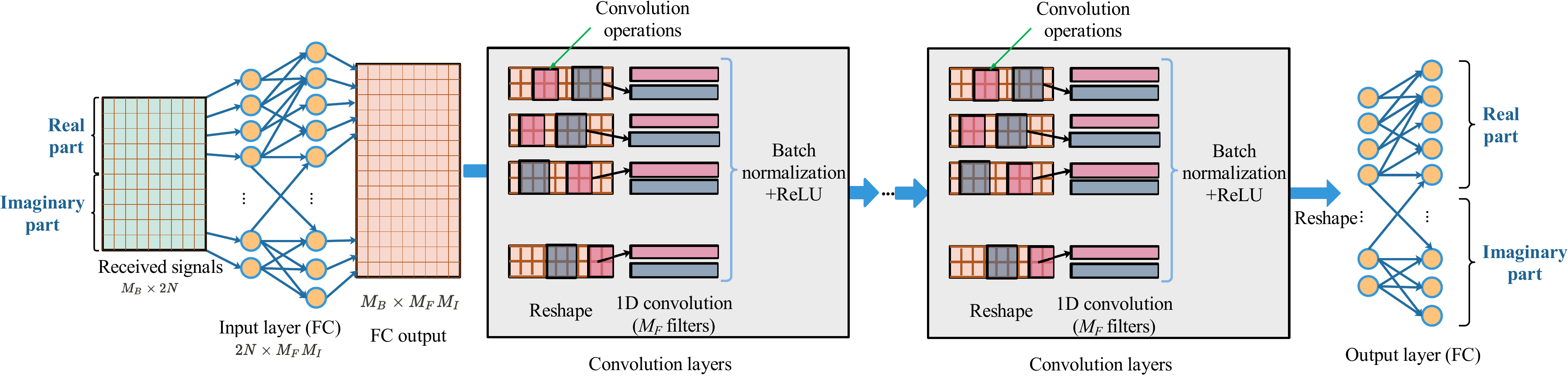}
	\caption{The network architecture of the proposed SDOA-Net.}
	\label{sysnet}
\end{figure*} 

\subsection{The MUSIC-Based Estimated Methods} 
In the super-resolution estimation method, the MUSIC-based methods can perform better by using noise and signal subspaces. For single-snapshot spectral estimation, ref~\cite{liao_music_2016} proposes a MUSIC-based method. A Hankel matrix is obtained from the received signal $\boldsymbol{r}$ as
\begin{align}
	\boldsymbol{R}=\text{Hankel}(\boldsymbol{r})\begin{bmatrix}
		r_0 & r_1 & \dots & r_{N-L}\\
		r_1 & r_2 & \dots & r_{N-L+1}\\
		\vdots & \vdots & \ddots & \vdots\\
		r_{L-1} & r_{L} & \dots & r_{N-1}
	\end{bmatrix},
\end{align}
where the received signal $\boldsymbol{r}$ is reshaped as a matrix $\boldsymbol{R}\in\mathbb{C}^{L\times N-L+1}$. Then, a singular value decomposition (SVD) is used as
\begin{align}
	[\boldsymbol{U}_1, \boldsymbol{U}_2] \boldsymbol{\Lambda}[\boldsymbol{V}_1,\boldsymbol{V}_2]=\text{SVD}\{\boldsymbol{R}\},
\end{align}
where $\boldsymbol{U}_2$ is corresponding to the small singular values and $\boldsymbol{\Lambda}$ is a diagonal matrix with the entries from the singular values. Finally, the spatial spectrum can be estimated as
\begin{align}
	g(\theta) = \frac{1}{\|\boldsymbol{a}^{\text{H}}(\theta)\boldsymbol{U}_2\|^2_2}.
\end{align} 

\section{The Proposed DOA Estimation Method}\label{sec4}

From the above sections about the existing DOA estimation methods, we can find that the DOAs are estimated by searching the peak values of the spatial spectrum. In this section, we will propose a DL-based super-resolution method for DOA estimation, and it is named a super-resolution DOA network, which contains more information and can be trained faster than the existing covariance matrix-based methods. 

\subsection{The Architecture of SDOA-Net}

%\begin{figure}
%	\centering
%	\includegraphics[width=2in]{./Figures/relu.pdf}
%	\caption{The ReLU function.}
%	\label{relu}
%\end{figure} 

The SDOA-Net architecture is shown in Fig.~\ref{sysnet}. First, the received signal in (\ref{eq1}) is rewritten as a vector with real and imaginary parts
\begin{align}
	\boldsymbol{y}(t)\triangleq [\mathcal{R}^{\text{T}}\{\boldsymbol{r}(t)\},\mathcal{I}^{\text{T}}\{\boldsymbol{r}(t)\}]^{\text{T}}\in\mathbb{R}^{2N\times 1},
\end{align}
where we have the received signal vector 
\begin{align}
	\boldsymbol{r}(t)=[r_0(t),r_1(t),\dots,r_{N-1}(t)]^\text{T}\in\mathbb{C}^{N\times 1}.
\end{align}
With the batch size being $M_B$, the  input signal is
\begin{align}
	\boldsymbol{Y}\triangleq [\boldsymbol{y}(0), \boldsymbol{y}(1),\dots,\boldsymbol{y}(M_B-1)]^{\text{T}},
\end{align} 
and the size is $M_B\times 2N$. 

Then, since the SDOA-Net is based on the convolution network, we use a full connection (FC) as the input layer with the output dimension being $M_FM_I$, where $M_F$ denotes the number of filters in the convolution layers and $M_I$ denotes the extension of the inner dimension. After the input layer, the dimension of the signal is $M_B\times M_FM_I$, and we reshape the signal as a tensor $f_1(\boldsymbol{Y})$ with the dimension $M_B \times M_F\times M_I$, where $f_1(\cdot)$ is an input layer function.

The tensor is passed to the convolution layers and the number of convolution layers is $M_{C}$. In each convolution layer, a one-dimensional (1D) convolution operation is realized with the kernel size being $M_F\times M_K$ and the padding operation is used to keep the size of the convolution output the same as that of the input. The output of the convolution operation is $M_B\times M_F \times M_I$. Then, the batch normalization is applied to the convolution output, and the normalization output is denoted as $f_3(f_2(f_1(\boldsymbol{Y})))$. The function $f_2(\cdot)$ denotes the convolution operation and $f_3(\cdot)$ is the batch normalization operation.
\begin{align}
f_3(x)=\frac{x-\mathcal{E}\{x\}}{\sqrt{\text{Var}\{x\}+\epsilon}},
\end{align}
where $\mathcal{E}\{x\}$ and $\text{Var}\{x\}$ are the mean and variance of $x$, respectively. $\epsilon$ is a value added to the denominator for numerical stability and can be set as $\epsilon=10^{-5}$. In each convolution layer, a ReLU function $f_4(\cdot)$ is applied to the output of the batch normalization and is defined as 
\begin{align}
	f_4(x)\triangleq \max(0,x).
\end{align}
After the convolution layers, an FC layer is used as an output layer with the input and output sizes being $M_B\times M_I M_F$ and $M_B\times 2N$, respectively. The operation in the output layer is denoted as $f_5(\cdot)$.

\begin{figure}
	\centering
	\includegraphics[width=1.2in]{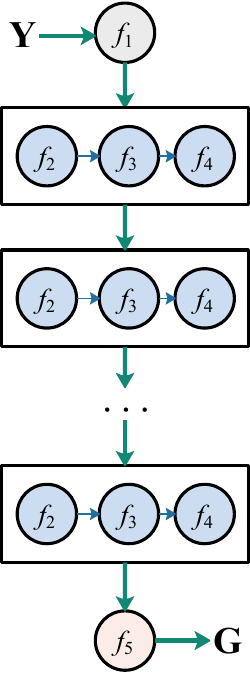}
	\caption{The flowchart of the functional operations.}
	\label{funcflow}
\end{figure}

Finally, as shown in Fig.~\ref{funcflow}, we can obtain the output of the SDOA-Net as
\begin{align}
\boldsymbol{G}=f_5(f_4(f_3(f_2(f_4(\dots f_4(\dots f_2(f_1(\boldsymbol{Y}))\dots)\dots ))))),
\end{align}
where we have 
\begin{align}
	\boldsymbol{G}\triangleq \begin{bmatrix}
\boldsymbol{g}(0),\boldsymbol{g}(1),\dots,\boldsymbol{g}(M_B-1)
\end{bmatrix}\in\mathbb{R}^{2N\times M_B}.
\end{align}

\begin{figure}
	\centering
	\includegraphics[width=3.5in]{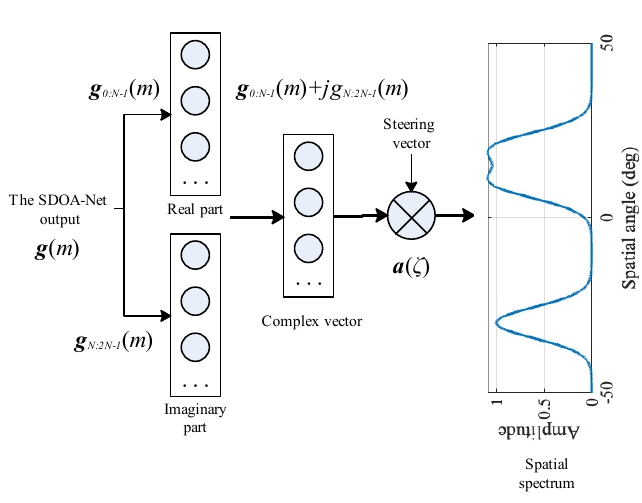}
	\caption{The flowchart to obtain the spatial spectrum from the network output.}
	\label{getsp}
\end{figure} 

As shown in Fig.~\ref{getsp}, the corresponding complex vector can be obtained from the network output $\boldsymbol{g}(m)$ ($m=0,1,\dots,M_B-1$) as
\begin{align}
\boldsymbol{z}(m)\triangleq \boldsymbol{g}_{0:N-1}(m)  +j \boldsymbol{g}_{N:2N-1}(m),
\end{align}
where $ \boldsymbol{g}_{0:N-1}(m) $ denotes a sub-vector of $\boldsymbol{g}(m)$ with the index  from $0$ to $N-1$, and $ \boldsymbol{g}_{N:2N-1}(m) $ denotes that from $N$ to $2N-1$. With the output  $\boldsymbol{z}$ of SDOA-Net, the spatial spectrum can be estimated by 
\begin{align}\label{eq20}
f_{\text{sp}}(\zeta) = |\boldsymbol{a}^{\text{H}}(\zeta)\boldsymbol{z}|^2,
\end{align}
where $\zeta$ is chosen based on the detection area, such as from $-\rm{\pi}/2$ to $\rm{\pi}/2$.

The SDOA-Net proposed in this study introduces a novel approach compared to existing methods. Unlike previous approaches, our network takes raw sampled data as input and utilizes convolution layers to extract features from these raw data. By using the raw data, which contain all the information of the received signals, we obtain a vector as the network's output. This output vector is distinct from the DOA results or the spatial spectrum used by existing methods. In particular, the size of the output vector matches the number of antennas in the array, resulting in a lower dimension compared to networks that output the spectrum. Consequently, training time can be significantly reduced. Furthermore, DOAs can be obtained by finding the peak values of $f_{\text{sp}}(\zeta)$ in (\ref{eq20}), which can avoid the problem of adopting the determined number of received signals in networks using DOA values as output.

\subsection{The Training Approach}
\begin{figure}
	\centering
	\includegraphics[width=3.7in]{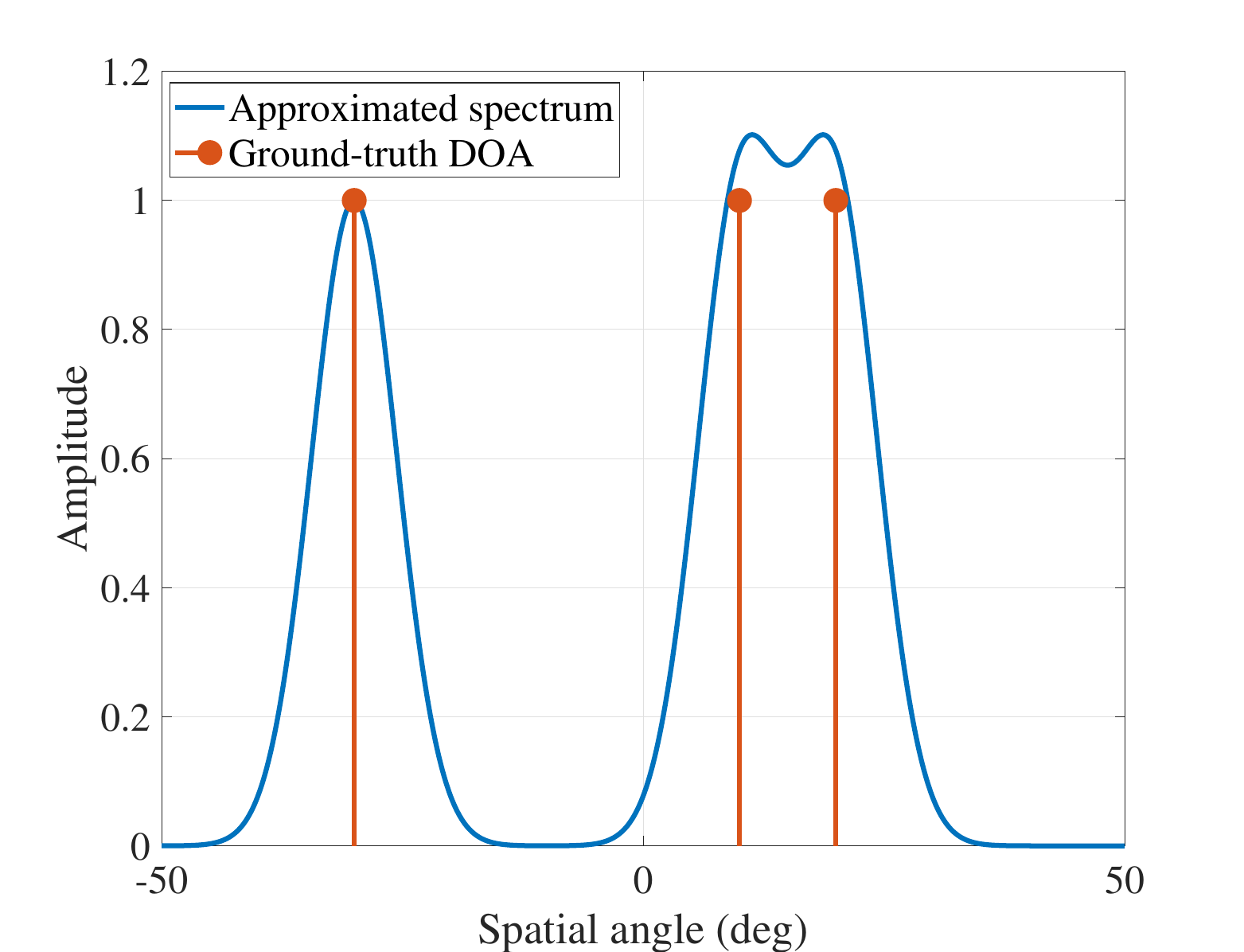}
	\caption{The refereed spatial spectrum approximated by the Gaussian functions.}
	\label{refsp}
\end{figure}

To train the SDOA-Net, the spatial spectrum $f_{\text{sp}}(\zeta)$ is obtained from (\ref{eq20}) and the refereed spectrum is given as follows
\begin{align}
f_{\text{ref}}(\zeta) = \sum_{k=0}^{K-1} A_k \text{e}^{-\frac{(\zeta-\theta_k)^2}{\sigma_{\text{G}}^2}},
\end{align}
where we use Gaussian functions to approximate the spatial spectrum. $A_k$ denotes the spectrum value, and $\sigma_{\text{G}}$ is the standard deviation of the Gaussian function. In this paper, we set the value of $\sigma_{\text{G}}$ as 
\begin{align}
	\sigma_{\text{G}}=\bar{\sigma}_{\text{G}}/N.
\end{align}
An example of the referenced spatial spectrum approximated by the Gaussian functions is shown in Fig.~\ref{refsp}, where we use $16$ antennas, $\bar{\sigma}_{\text{G}}=100$, and the ground-truth DOAs are $\ang{-30}$, $\ang{10}$, and $\ang{20}$. The width of the $3$ dB spectrum is about $\ang{10.4}$.

With the refereed spectrum, the loss function is defined as
\begin{align} \label{eq22}
f_{\text{loss}}(\boldsymbol{\zeta}) =\frac{1}{\Omega} \left\|f_{\text{ref}}(\boldsymbol{\zeta})-f_{\text{sp}}(\boldsymbol{\zeta})\right\|^2_2,
\end{align}
where $f_{\text{ref}}(\boldsymbol{\zeta})\in\mathbb{R}^{\Omega\times 1}$ and $f_{\text{sp}}(\boldsymbol{\zeta})\in\mathbb{R}^{\Omega\times 1}$ are vectors with the $\omega$-th ($\omega=0,1,\dots, \Omega-1$) entry being $f_{\text{ref}}(\zeta_\omega)$  and $f_{\text{sp}}(\zeta_\omega)$, respectively. We define 
\begin{align}
	\boldsymbol{\zeta}\triangleq \begin{bmatrix}
\zeta_0,\dots, \zeta_{\Omega-1}
\end{bmatrix}^{\text{T}},
\end{align}
where $\Omega$ is the number of the discretized spatial angles. The SDOA-Net is trained to minimize the loss function $f_{\text{loss}}(\boldsymbol{\zeta})$ in (\ref{eq22}) by updating the network coefficients.

\begin{figure}
	\centering
	\includegraphics[width=2.5in]{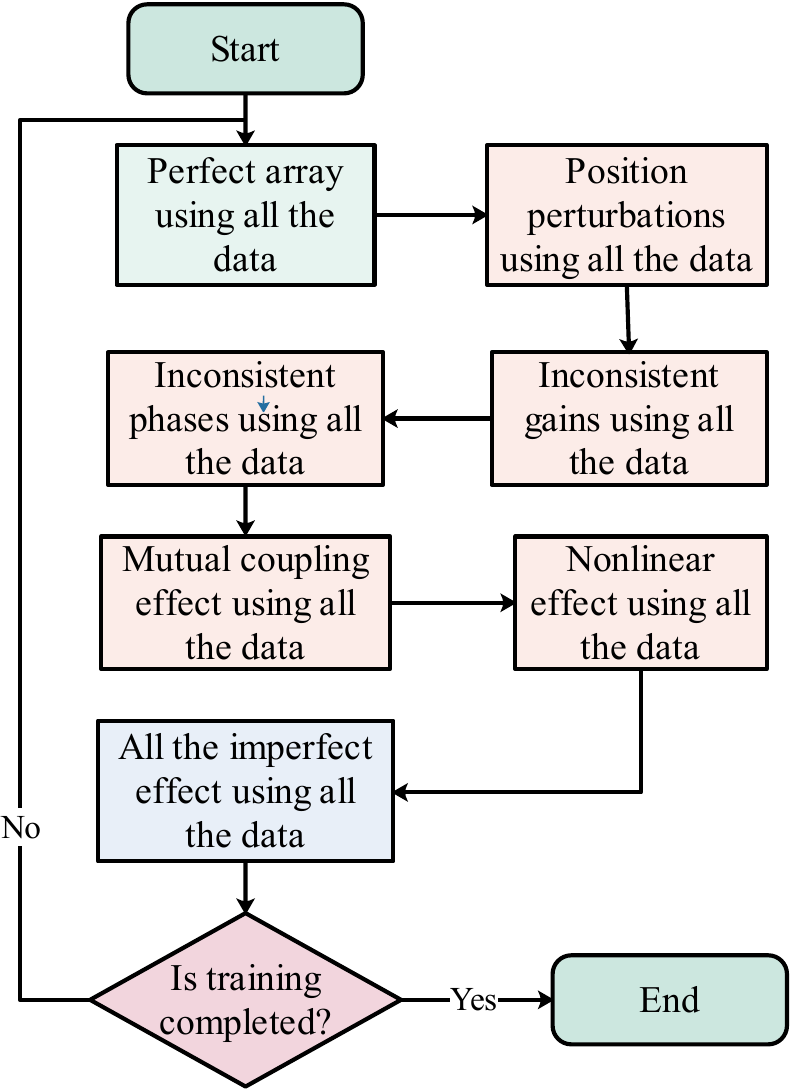}
	\caption{The training procedure for the SDOA-Net.}
	\label{trainflow}
\end{figure} 

For the practical system, the mutual coupling effect, the nonlinear effect, the inconsistent phases, the inconsistent gains, and the position perturbations are considered in this paper. The training procedure is shown in Fig.~\ref{trainflow}, and the following steps can be used to train the SDOA-Net:
\begin{enumerate}
\item \textbf{Perfect array step:} The received signals using perfect array without the imperfect effect are used during the training procedure;
\item \textbf{Position perturbation step:} The received signals with position perturbation are used. The position perturbation is generated by a Gaussian distribution with the mean being $0$ and the standard deviation $\sigma_{\text{per}}$ selected by a uniform distribution $\sigma_{\text{per}}\in [0,\sigma_{\text{max\_per}}]$. The parameter $\sigma_{\text{max\_per}}$ can be specified in the simulation;
\item \textbf{Inconsistent gains step:} The inconsistent gains are considered in this step. Similarly, the inconsistent gains are generated by a zero-mean Gaussian distribution with the standard deviation $\sigma_{\text{gain}}$ being $\sigma_{\text{gain}}\in [0, \sigma_{\text{max\_gain}}]$, where $\sigma_{\text{max\_gain}}$ is specified in the simulation;
\item \textbf{Inconsistent phases step:} The inconsistent phases are also generated by a zero-mean Gaussian distribution with the  standard deviation $\sigma_{\text{phase}}$ being $\sigma_{\text{phase}}\in [0, \sigma_{\text{max\_phase}}]$, where $\sigma_{\text{max\_phase}}$ is specified in the simulation;
\item \textbf{Mutual coupling effect step:} The mutual coupling effect is described by a matrix $\boldsymbol{B}$ with complex entries, and the diagonal entries are all ones. The entry at the $n$-th row and the $n'$-th column is denoted as 
\begin{align}
	B_{n,n'}=|B_{n,n'}|\text{e}^{j\psi_{n,n'}},
\end{align}
and $|B_{n,n'}|$  is determined by a uniform distribution $|B_{n,n'}|\in[0,\sigma^{|n-n'|}_{\text{mc}}]$ with $n\neq n'$.  The phase $\psi_{n,n'}$ follows a uniform distribution $\psi_{n,n'}\in[0,2\rm{\pi})$. The parameter $\sigma_{\text{mc}}$ is specified in the simulation;
\item \textbf{Nonlinear effect step:} The nonlinear effect is described by a nonlinear function
\begin{align}
	f_{\text{nonlinear}}(x)=\tanh(x\sigma_{\text{nonlinear}}),
\end{align}
where $\sigma_{\text{nonlinear}}$ is specified in the simulation to control the nonlinear effect. The $\text{tanh}(\cdot)$ function is used as a nonlinear function. Usually, we can also use other types of the activation function, such as ReLU, leaky ReLU, sigmoid, etc., and the choice of activation function will not have a big impact on the performance of the DOA estimation;
\item \textbf{All the imperfect effect step:} We consider all the imperfect effects to train the network.
\end{enumerate}
The network is trained to the next step after all the data is used in the current step. For example, when we use all the data in the inconsistent phases step, we go to the next step of the mutual coupling effect. After training SDOA-Net in sequence according to the above steps, we start over from the first step to train the network again until the maximum number of training procedures.   

\section{Simulation Results} \label{sec5}
In this section, the DOA estimation performance of the proposed SDOA-Net using a practical array is evaluated through simulation. The simulations are conducted on a computer with MATLAB R2020b, equipped with an Intel Core i5 @ 2.9 GHz processor and 8 GB LPDDR3 @ 2133 MHz. The SDOA-Net source code, including the training codes and a pre-trained network, is available at https://github.com/chenpengseu/SDOA-Net.git. SDOA-Net is based on PyTorch~1.4 and Python~3.7. The simulation parameters are given in Table~\ref{table1}.  We use $N=16$ antennas to receive the signals and the SDOA-Net to estimate the DOA, where the number of signals is $K=3$. Moreover, the hyperparameters for the imperfect array are also given in Table~\ref{table1}. The estimation performance is measured by the root mean square error (RMSE)
\begin{align}
	\text{RMSE} = \sqrt{\frac{1}{N_{\text{sim}}K}\left\|
		\hat{\boldsymbol{\theta}}-\boldsymbol{\theta}\right\|^2_2},
\end{align}
where $N_{\text{sim}}$ is the number of simulations,  $\hat{\boldsymbol{\theta}}$ is the estimated DOA vector, and $\boldsymbol{\theta}$ is the ground-truth DOA vector.

\begin{table}% [!t] 
	\renewcommand{\arraystretch}{1.3}
	\caption{Simulation parameters}
	\label{table1}
	\centering
	\begin{tabular}{cc}
		\hline
		\textbf{Parameter}                 & \textbf{Value}                                                   \\
		\hline
		The standard deviation in the Gaussian function  & $\bar{\sigma}_{\text{G}}=100$  \\
		The batch size         & $64$                                                           \\
		The number of convolution layers & $6$                                                    \\
		The number of filters in the convolution layer               & $2$                                                            \\
		The kernel size in the convolution layer                & $3$                                          \\
		The learning rate          & $5\times 10^{-4}$ \\
		The number of antennas  $N$       & $16$ \\
		The number of targets   $K$      & $3$ \\
		The distance between adjacent antennas & $0.5\lambda$ \\
		\tabincell{c}{The maximum standard deviation of\\ position perturbation ${\sigma}_{\text{max\_per}}$} 
		  & $0.15$  \\
		\tabincell{c}{The maximum standard deviation of\\ inconsistent gain ${\sigma}_{\text{max\_gain}}$}  & $0.5$  \\
		\tabincell{c}{The maximum standard deviation of \\inconsistent phase ${\sigma}_{\text{max\_phase}}$}  & $0.2$  \\
		The maximum mutual coupling effect ${\sigma}_{\text{mc}}$  & $0.06$  \\
		The nonlinear effect ${\sigma}_{\text{nonlinear}}$ & $1.0$  \\
		\hline
	\end{tabular}
\end{table}
    
\begin{figure}
	\centering
	\includegraphics[width=3.7in]{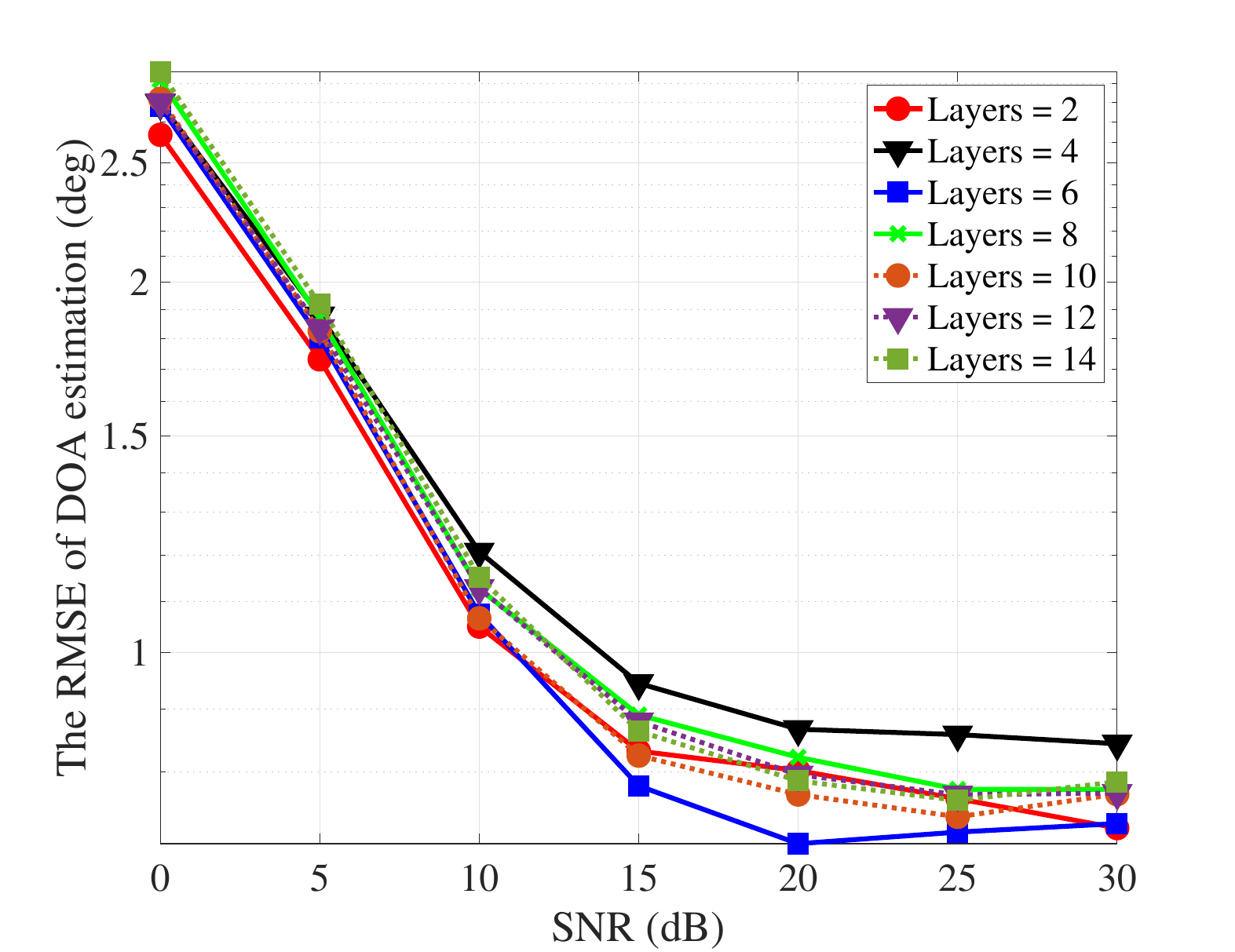}
	\caption{The DOA estimation performance with different numbers of layers.}
	\label{layer}
\end{figure}

First, the proposed SDOA-Net contains convolution layers and each convolution layer has convolution, batch normalization, and ReLU active function operations. In SDOA-Net, some important hyperparameters must be considered for a better DOA estimation. The first hyperparameter is the number of 1D convolution layers. In Fig.~\ref{layer}, we show the performance of the DOA estimation with different numbers of convolution layers. As shown in this figure, when the number of convolutions is $6$, a better estimation performance is achieved, so we use $6$ convolution layers in the following simulations.

\begin{figure}
	\centering
	\includegraphics[width=3.7in]{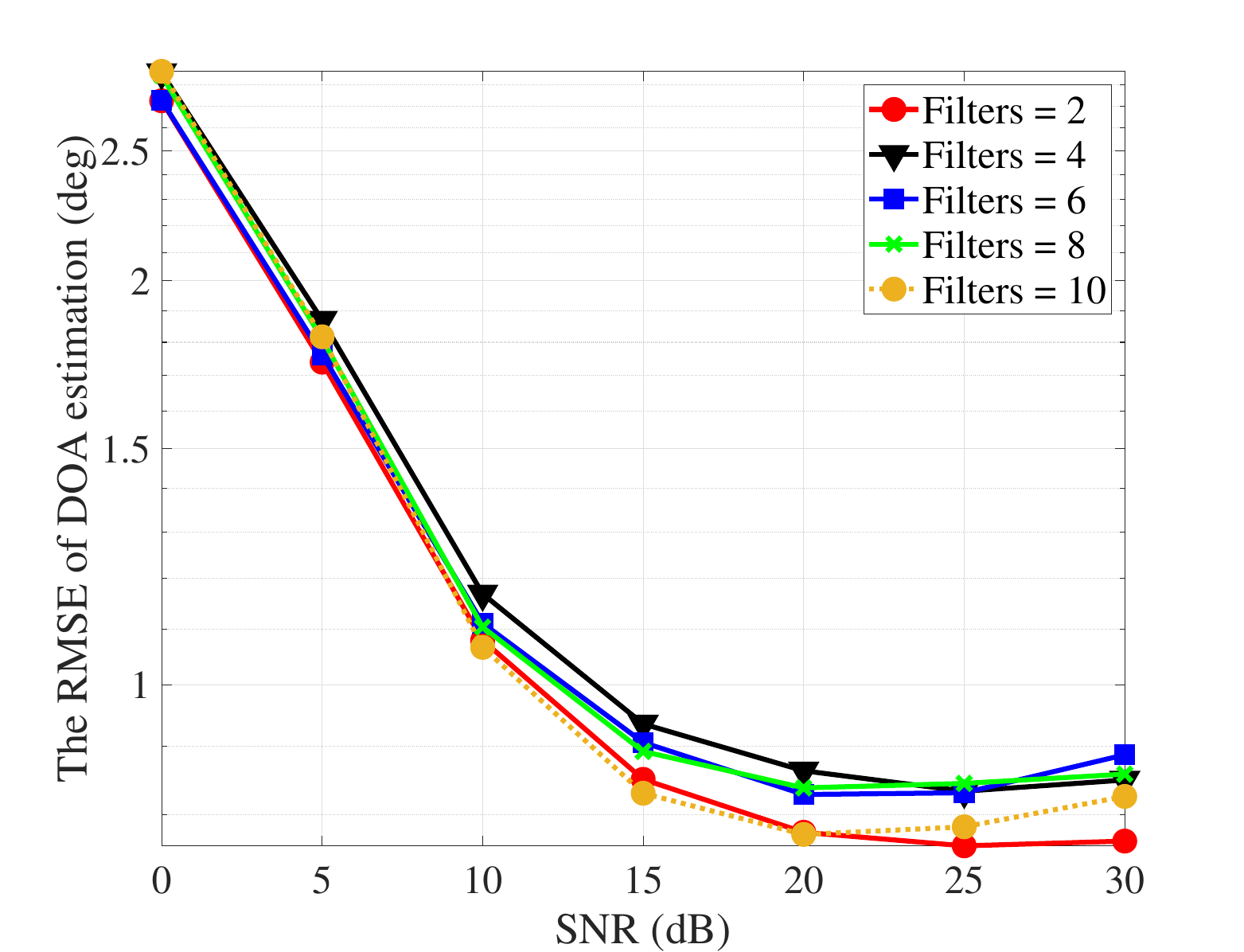}
	\caption{The DOA estimation performance with different numbers of filters.}
	\label{filter}
\end{figure}

Then, we compare the DOA estimation performance among the networks using different numbers $M_F$ of filters that are used in the convolution layers. As shown in Fig.~\ref{filter}, for the consideration of both the estimation performance and the network complexity, better performance is achieved with $M_F=2$, so we will use $2$ filters in the following simulations.  Note that selecting optimal values for individual parameters does not necessarily ensure that the network configuration will attain the global optimum. Nonetheless, by comparing the network's performance across various parameter settings, we are able to assess the influence of distinct parameters on the network's overall performance. This analysis facilitates informed the selection of network parameters.

\begin{figure}
	\centering
	\includegraphics[width=3.7in]{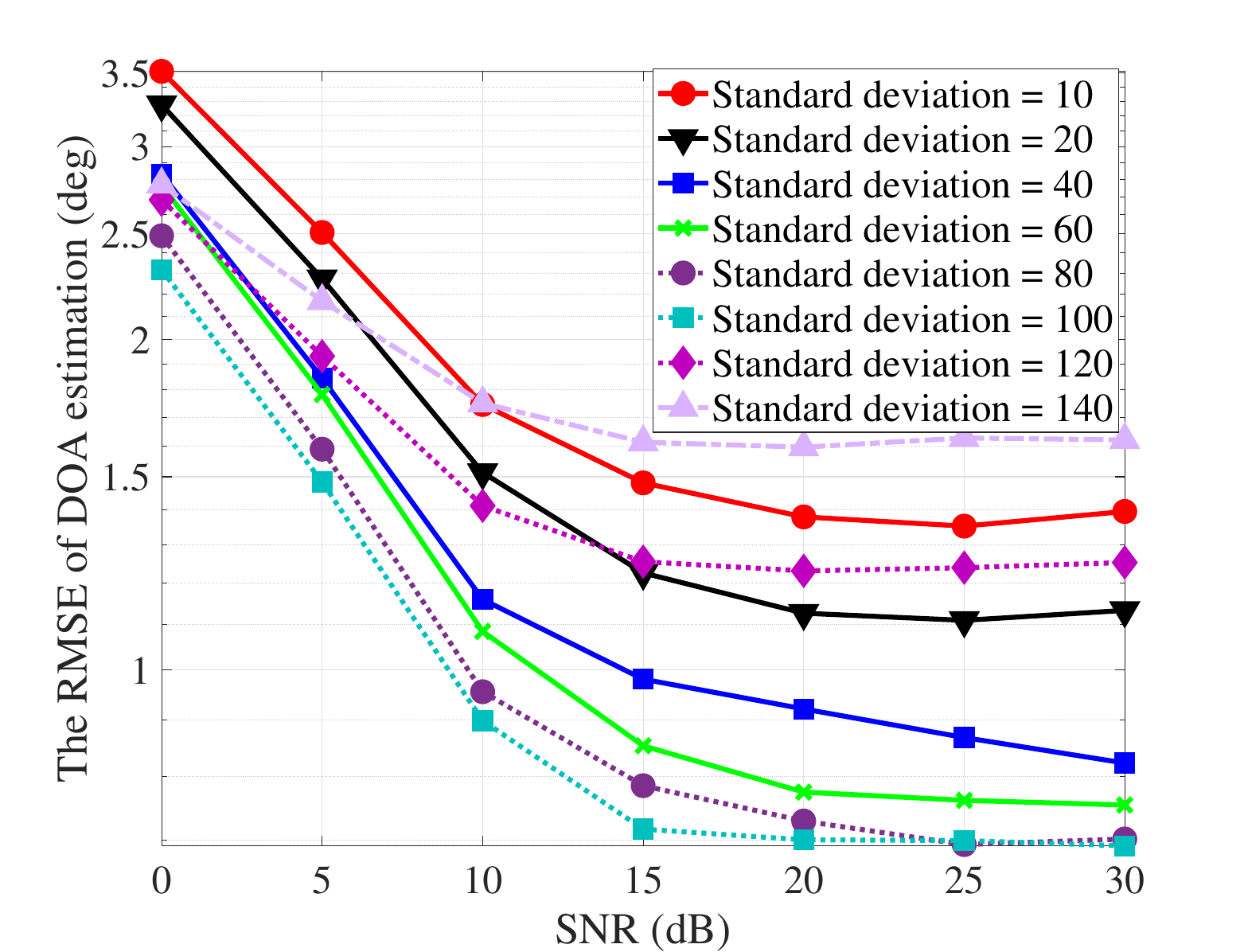}
	\caption{The DOA estimation performance with different standard deviations $\bar{\sigma}_{\text{G}}$.}
	\label{std}
\end{figure} 

In the procedure of training the SDOA-Net, the referred spatial spectrum is used to measure the loss function, where we use the Gaussian functions to approximate the spatial spectrum. Hence, the standard deviation $\bar{\sigma}_{\text{G}}$ in the Gaussian function is important to approximating the spatial spectrum. We show the performance of DOA estimation with different standard deviations $\bar{\sigma}_{\text{G}}$ in Fig.~\ref{std}. When the standard deviation $\bar{\sigma}_{\text{G}}$ is $100$, a better performance of DOA estimation is achieved, so we will use $\bar{\sigma}_{\text{G}}=100$ in the following simulation. 

\begin{figure}
	\centering
	\includegraphics[width=3.8in]{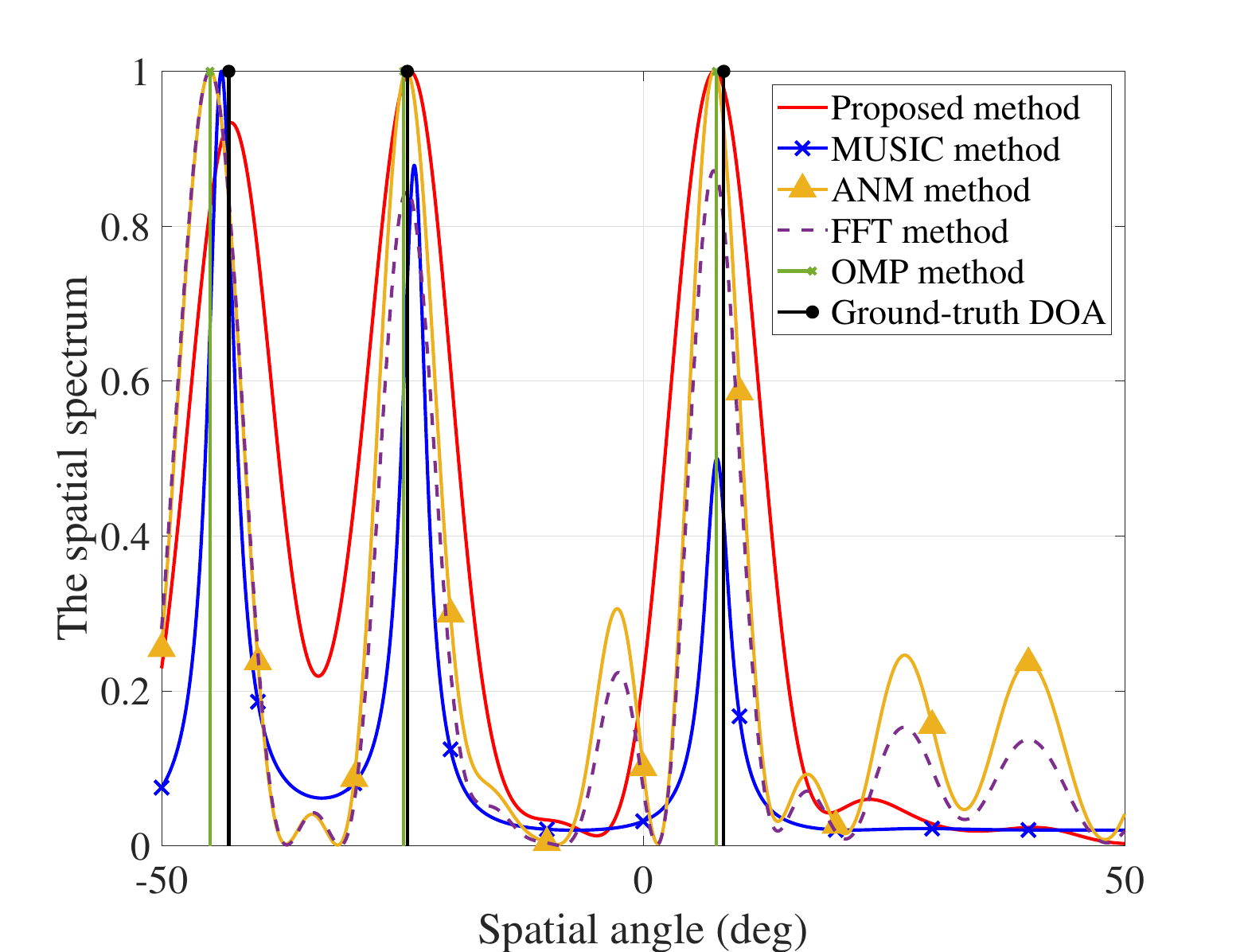}
	\caption{The spatial spectrum  compared with the existing methods.}
	\label{spectrum}
\end{figure} 

Next, based on the above SDOA-Net parameters, the estimated spatial spectrum is shown in Fig.~\ref{spectrum} for the DOA estimation and is also compared with the following existing methods:
\begin{itemize}
	\item \textbf{MUSIC method~\cite{liao_music_2016}:} The conventional MUSIC method estimates the covariance matrix based on multiple snapshots and employs eigenvalue decomposition to estimate DOA. To make a fair comparison, we adopt the snapshot-based MUSIC algorithm proposed in~\cite{liao_music_2016} that utilizes a Hankel data matrix and Vandermonde decomposition in the MUSIC method.
	\item \textbf{ANM method~\cite{govinda_raj_single_2019,wei_gridless_2020,zai_yang_enhancing_2016}:} ANM-based methods have been introduced for DOA estimation, which can take advantage of the sparsity of the targets in the spatial domain. In contrast to current CS-based methods, which involve discretizing the spatial domain into grids and using a dictionary matrix for sparse reconstruction, such as those proposed in~\cite{z_yang_orthonormal_2011,z_tan_joint_2014,g_yu_statistical_2011}, ANM methods estimate DOA in the continuous domain. This approach can overcome the \emph{off-grid} problem caused by discrete methods.
 
	\item \textbf{FFT method:} The FFT method is widely used in practical systems with low computational complexity. However, the resolution of the FFT method is unsatisfactory but robust to the imperfect array;
	\item \textbf{OMP method~\cite{aghababaiyan_high-precision_2020,lin_single_2021,chen_source_2019}:} The 
	  orthogonal matching pursuit (OMP) method is a CS-based method using the discretized spatial angles and has relatively low computational complexity. Hence, it has been widely used in sparse reconstruction problems.
\end{itemize} 
As shown in Fig.~\ref{spectrum}, the spatial spectrum estimated by the proposed SDOA-Net performs better than the MUSIC, ANM, FFT and OMP methods. Additionally, the proposed method is based on the convolution network and has low computational complexity than the ANM and MUSIC methods. The computational complexity of the proposed network is $\mathcal{O}(N^2)$, and the computational complexities of ANM and MUSIC are $\mathcal{O}((N+1)^{6.5})$ and $\mathcal{O}(N^3)$, respectively. Therefore, the proposed SDOA-Net is efficient in the DOA estimation problem.

\begin{figure}
	\centering
	\includegraphics[width=3.7in]{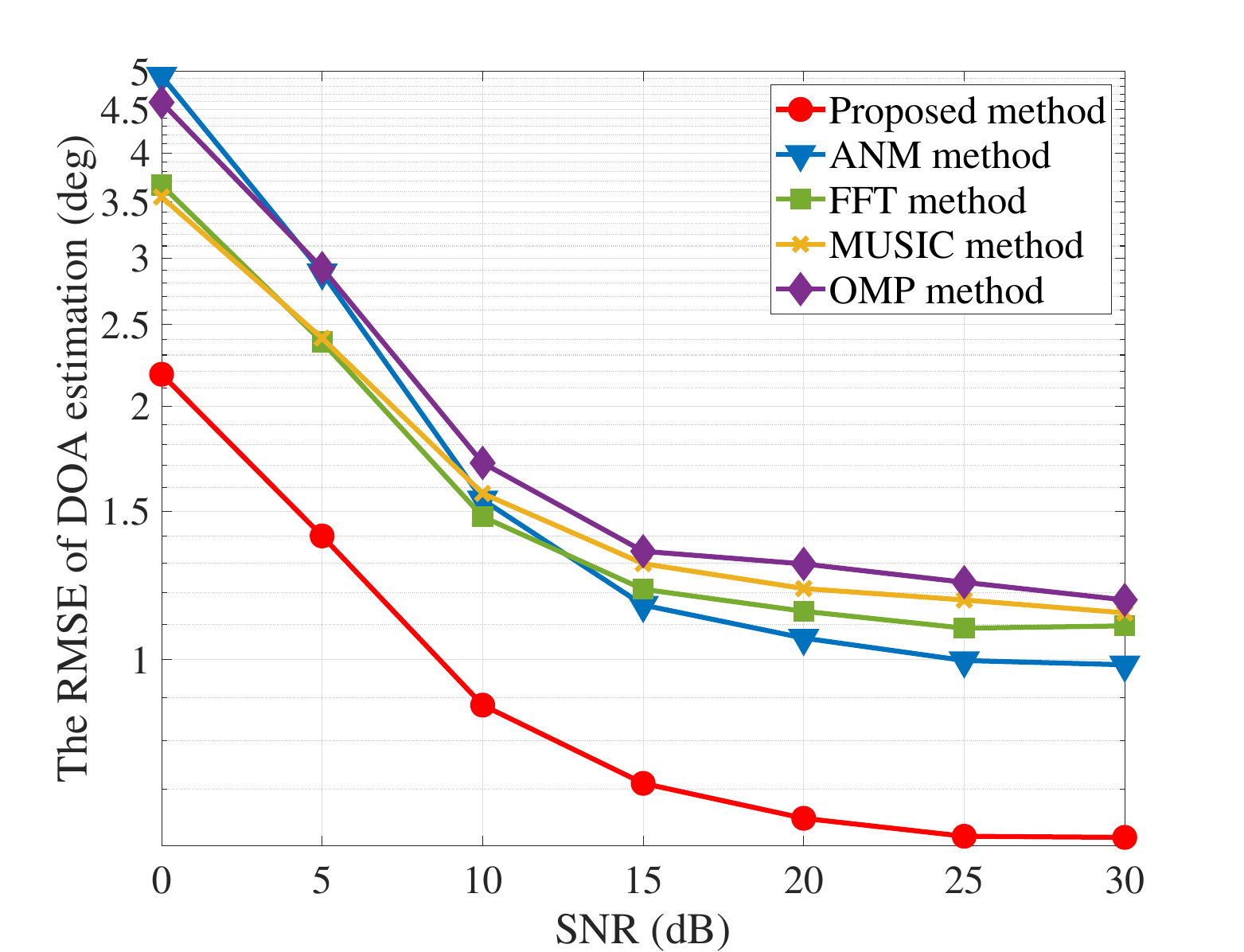}
	\caption{The DOA estimation performance with different SNRs.}
	\label{snr}
\end{figure} 

Next, the performance of the DOA estimation under different SNRs is shown in Fig.~\ref{snr}, where the SNR ranges from $0$ dB to $30$ dB. This figure shows that the proposed method achieves a better estimation performance in the scenario with an imperfect array than the method using the ANM, FFT, MUSIC, and OMP methods. For the SNR being $10$~dB, the RMSE of the proposed SDOA-Net is about $\ang{0.70}$ and that of the ANM method is about $\ang{1.15}$, so the RMSE improvement is about $39.13\%$. Furthermore, when the SNR is $7.5$~dB, the RMSE of the proposed SDOA-Net method is the same as that of the ANM method with the SNR being $15$~dB, so the improvement in the SNR is about $7.5$~dB.

\begin{figure}
	\centering
	\includegraphics[width=3.7in]{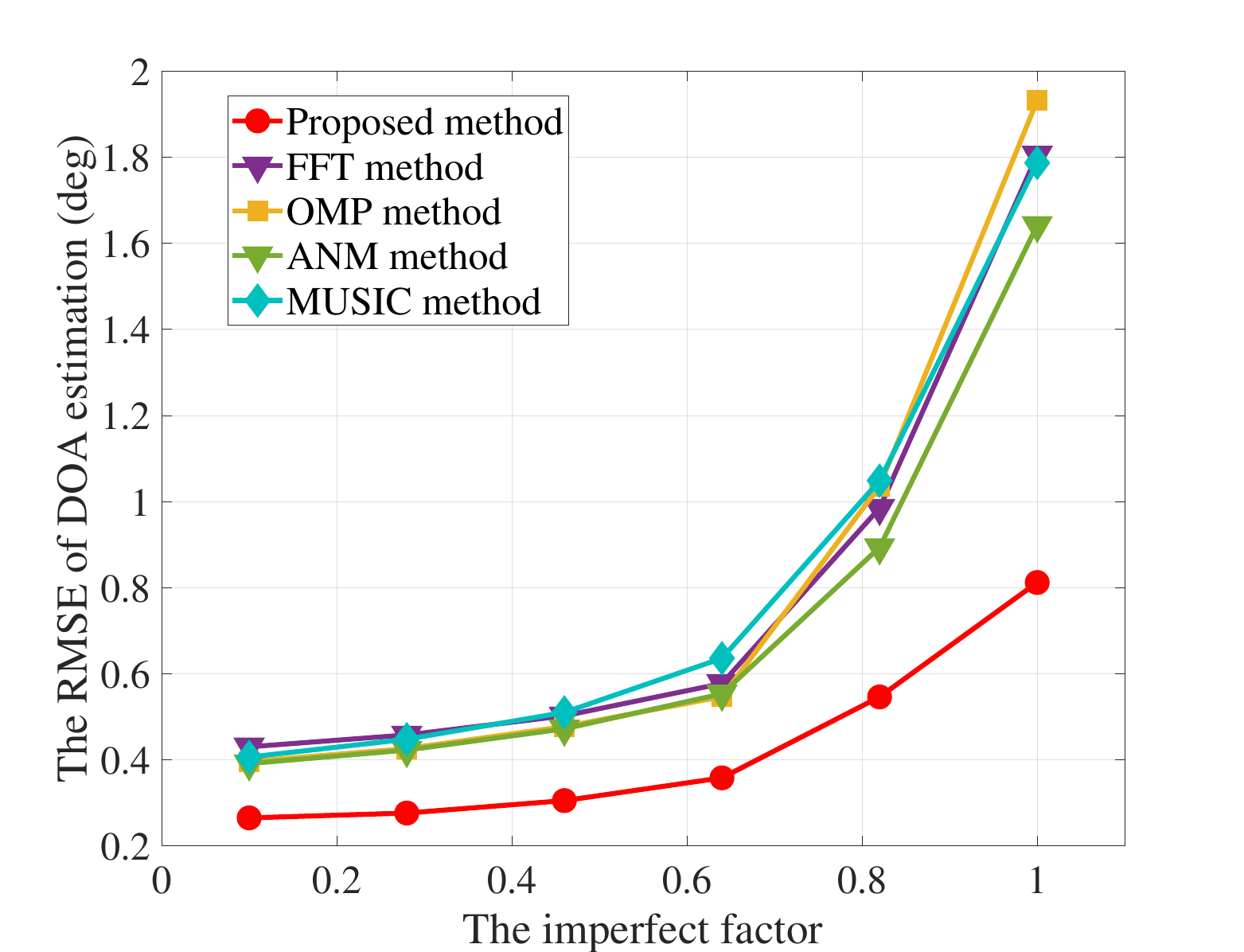}
	\caption{The DOA estimation performance with different imperfect factors.}
	\label{imperfect}
\end{figure} 

We use an imperfect factor to measure the imperfect effect, defined as $\xi$. With the imperfect factor $\xi$, the imperfect parameters for position perturbation, inconsistent gain, inconsistent phase, mutual coupling effect and non-linear effect will be $\xi\sigma_{\text{max\_per}}$, $\xi\sigma_{\text{max\_gain}}$, $\xi\sigma_{\text{max\_phase}}$, $\xi\sigma_{\text{mc}}$, and $\xi\sigma_{\text{nonlinear}}$, respectively. For example,  as presented in Table~\ref{table1}, the maximum standard deviation of position perturbation is denoted as  ${\sigma}_{\text{max\_per}}=0.15$.Given an imperfection factor of  $\xi=0.5$, the standard deviation of position perturbation for the simulations is adjusted to $\xi {\sigma}_{\text{max\_per}}=0.5\times 0.15=0.075$. To evaluate the DOA estimation performance across various scenarios, we vary the imperfection factor from $0.1$ to $1.0$.
 
Fig.~\ref{imperfect} illustrates the performance of DOA estimation when considering various imperfect factors. The SDOA-Net method, proposed in this study, outperforms the compared methods in terms of accuracy of estimation. Moreover, the proposed method demonstrates superior performance in scenarios with higher imperfect factors, indicating its robustness against the negative impact of imperfections.

\begin{figure}
	\centering
	\includegraphics[width=3.7in]{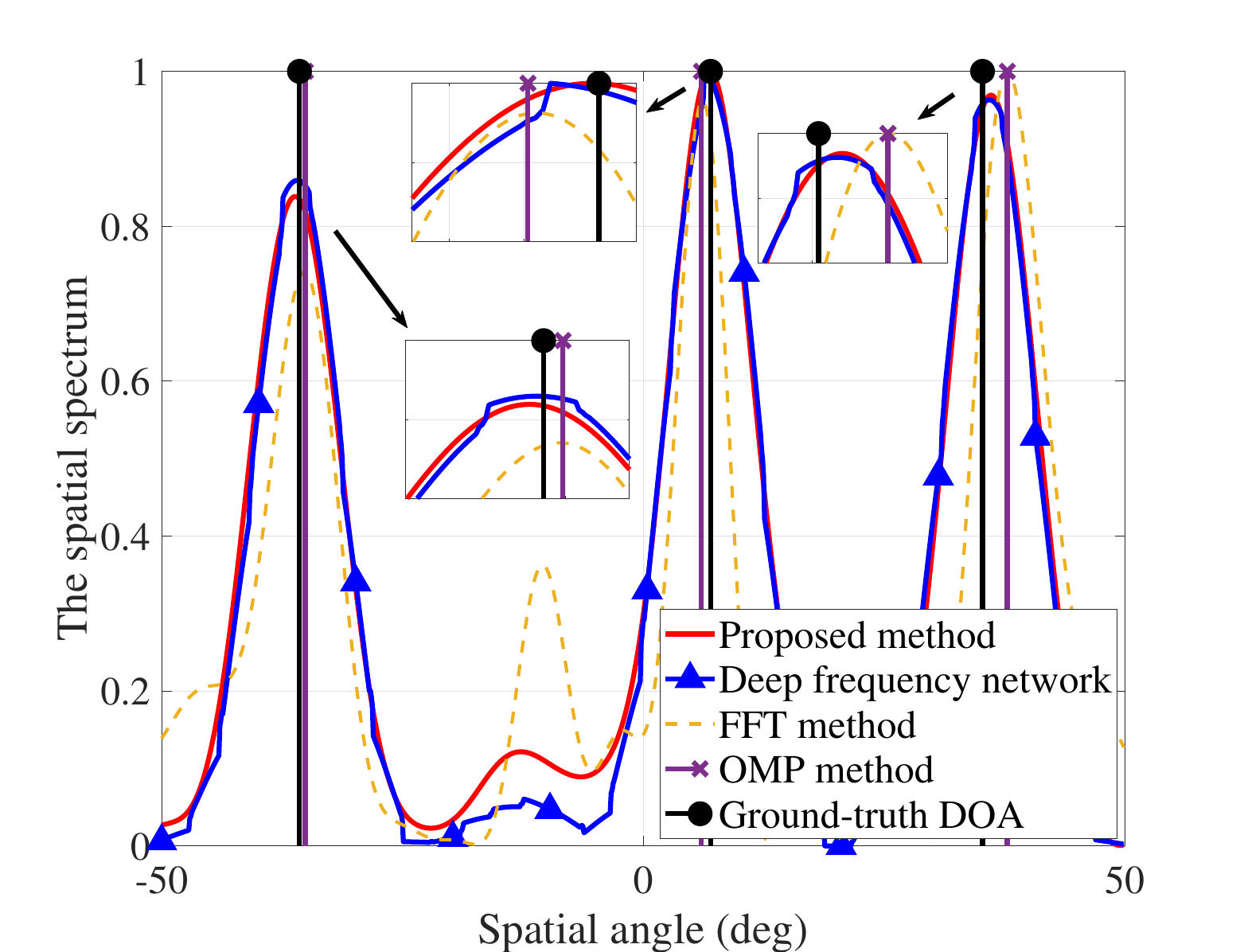}
	\caption{The spatial spectrum  compared with the existing deep learning-based method.}
	\label{dlsp}
\end{figure} 

In addition, ref.~\cite{izacard2019data} also introduces a DL-based approach to estimate the DOA, known as the deep frequency network. This method utilizes the network output as the spectrum. Fig.~\ref{dlsp} illustrates the spatial spectrum of both the proposed SDOA-Net and the deep frequency network. It can be observed that the estimated spectrum from the deep frequency network is less smooth compared to that of the proposed SDOA-Net, since the output of the deep frequency network is the spatial spectrum. Consequently, SDOA-Net demonstrates superior DOA estimation performance compared to the deep frequency network. 

\begin{figure}
	\centering
	\includegraphics[width=3.7in]{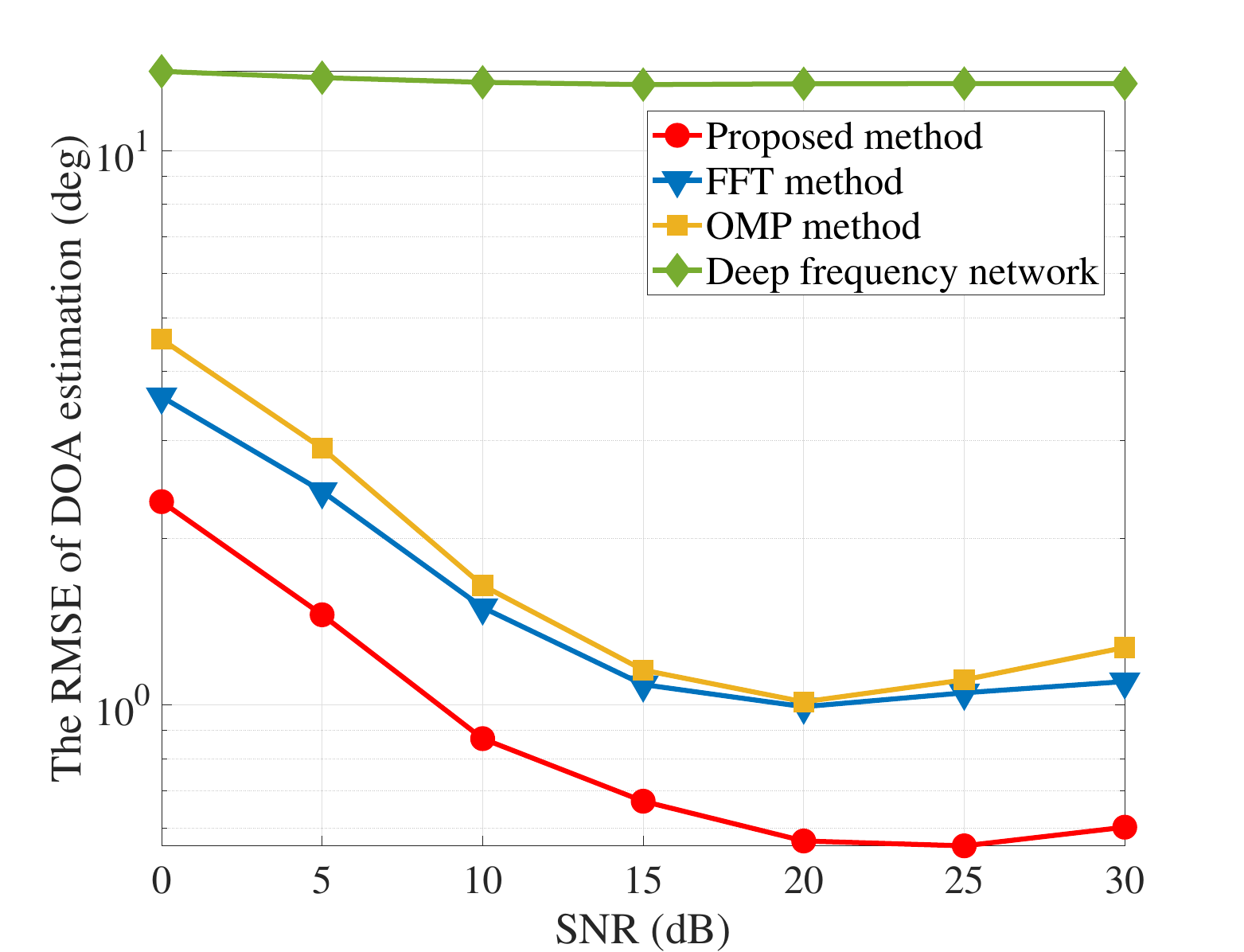}
	\caption{The DOA estimation performance compared with the existing deep learning-based method.}
	\label{dlsnr}
\end{figure}

Fig.~\ref{dlsnr} illustrates the performance of DOA estimation at varying SNRs, ranging from $0$~dB to $30$~dB. Both the SDOA-Net and the deep frequency network employ the same training data set. From the DOA estimation performance, it can be found that the deep frequency network does not achieve a better DOA estimation performance, mainly because of the following two reasons: 1) The spatial spectrum obtained by  the deep frequency network is not smooth enough, which is easy to incorrectly select the peak value of the spatial spectrum and fail to obtain the DOA information; 2) The deep frequency network has high accuracy for the DOA estimation of the perfect array~\cite{izacard2019data}, but its efficacy considerably diminishes for the imperfect array due to the lack of specific optimization for robustness.  Consequently, when compared to traditional model-based approaches such as FFT and OMP methods, the deep frequency network exhibits inferior performance. The performance of the proposed SDOA-Net surpasses that of existing methods such as the FFT method, the OMP method, and the deep frequency network. Since the proposed SDOA-Net is tailored for imperfect array  and generates a smoother output spatial spectrum, which simplifies the process of accurately identifying the spectral peaks during the peak-search stage. Consequently, this enhancement enables the proposed SDOA-Net to achieve the improved DOA estimation performance for the imperfect array.

\section{Conclusions}\label{sec6}
The problem of estimating the DOA has been studied in the context of an imperfect array. A system model has been developed to account for various factors such as antenna position perturbations, inconsistent gains and phases, mutual coupling effect, and nonlinear effect. To address this problem, a novel method, called SDOA-Net, has been introduced. Unlike existing approaches, SDOA-Net takes raw sampled signals as input and produces a vector that can be used to estimate the spatial spectrum. By utilizing convolution layers, SDOA-Net achieves faster convergence in training compared to other deep learning-based methods. The simulation results demonstrate the advantages of SDOA-Net in DOA estimation using a practical array. For an SNR of $10$ dB, the RMSE in DOA estimation achieved through the proposed method exhibits a $39.13\%$ improvement over the performance of the ANM method. In the future, further research will focus on the theoretical analysis of SDOA-Net's performance in DOA estimation. 
  
\bibliographystyle{IEEEtran}
\bibliography{IEEEabrv.bib,References.bib}

\begin{thebibliography}{10}
\providecommand{\url}[1]{#1}
\csname url@samestyle\endcsname
\providecommand{\newblock}{\relax}
\providecommand{\bibinfo}[2]{#2}
\providecommand{\BIBentrySTDinterwordspacing}{\spaceskip=0pt\relax}
\providecommand{\BIBentryALTinterwordstretchfactor}{4}
\providecommand{\BIBentryALTinterwordspacing}{\spaceskip=\fontdimen2\font plus
\BIBentryALTinterwordstretchfactor\fontdimen3\font minus
  \fontdimen4\font\relax}
\providecommand{\BIBforeignlanguage}[2]{{%
\expandafter\ifx\csname l@#1\endcsname\relax
\typeout{** WARNING: IEEEtran.bst: No hyphenation pattern has been}%
\typeout{** loaded for the language `#1'. Using the pattern for}%
\typeout{** the default language instead.}%
\else
\language=\csname l@#1\endcsname
\fi
#2}}
\providecommand{\BIBdecl}{\relax}
\BIBdecl

\bibitem{10005294}
Q.~Tian and R.~Cai, ``A low-complexity {DOA} estimation algorithm for
  distributed source localization,'' \emph{{IEEE} Trans. Instrum. Meas.},
  vol.~72, pp. 1--4, 2023.

\bibitem{zhang_overview_2021}
J.~A. Zhang, F.~Liu, C.~Masouros, R.~W. Heath, Z.~Feng, L.~Zheng, and
  A.~Petropulu, ``An overview of signal processing techniques for joint
  communication and radar sensing,'' \emph{IEEE Journal of Selected Topics in
  Signal Processing}, vol.~15, no.~6, pp. 1295--1315, Nov. 2021.

\bibitem{9845366}
M.~Pan, P.~Liu, S.~Liu, W.~Qi, Y.~Huang, X.~You, X.~Jia, and X.~Li, ``Efficient
  joint {DOA} and {TOA} estimation for indoor positioning with {5G} picocell
  base stations,'' \emph{{IEEE} Trans. Instrum. Meas.}, vol.~71, pp. 1--19,
  2022.

\bibitem{xu_rate-splitting_2021}
C.~Xu, B.~Clerckx, S.~Chen, Y.~Mao, and J.~Zhang, ``Rate-splitting multiple
  access for multi-antenna joint radar and communications,'' \emph{IEEE Journal
  of Selected Topics in Signal Processing}, vol.~15, no.~6, pp. 1332--1347,
  Nov. 2021.

\bibitem{zhu_combined_2018}
L.~Zhu, S.~Qiu, and Y.~Han, ``Combined constrained adaptive sum and difference
  beamforming in monopulse angle estimation,'' \emph{{IEEE} Antennas Wireless
  Propag. Lett.}, vol.~17, no.~12, pp. 2314--2318, Dec. 2018.

\bibitem{lin_fsf_2006}
J.~Lin, W.~Fang, Y.~Wang, and J.~Chen, ``{FSF} {MUSIC} for joint {DOA} and
  frequency estimation and its performance analysis,'' \emph{{IEEE} Trans.
  Signal Process.}, vol.~54, no.~12, pp. 4529--4542, Dec. 2006.

\bibitem{yan_low-complexity_2013}
F.~Yan, M.~Jin, and X.~Qiao, ``Low-complexity {DOA} estimation based on
  compressed {MUSIC} and its performance analysis,'' \emph{{IEEE} Trans. Signal
  Process.}, vol.~61, no.~8, pp. 1915--1930, Apr. 2013.

\bibitem{zhang_direction_2010}
X.~Zhang, L.~Xu, L.~Xu, and D.~Xu, ``Direction of departure ({DOD}) and
  direction of arrival ({DOA}) estimation in {MIMO} radar with
  reduced-dimension {MUSIC},'' \emph{{IEEE} Commun. Lett.}, vol.~14, no.~12,
  pp. 1161--1163, Dec. 2010.

\bibitem{kim_joint_2015}
S.~Kim, D.~Oh, and J.~Lee, ``Joint {DFT}-{ESPRIT} estimation for {TOA} and
  {DOA} in vehicle {FMCW} radars,'' \emph{{IEEE} Antennas Wireless Propag.
  Lett.}, vol.~14, pp. 1710--1713, 2015.

\bibitem{lin_time-frequency_2016}
J.~Lin, X.~Ma, S.~Yan, and C.~Hao, ``Time-frequency multi-invariance {ESPRIT}
  for {DOA} estimation,'' \emph{{IEEE} Antennas Wireless Propag. Lett.},
  vol.~15, pp. 770--773, 2016.

\bibitem{xiaofei_zhang_multi-invariance_2009}
X.~Zhang, X.~Gao, and D.~Xu, ``Multi-invariance {ESPRIT}-based blind {DOA}
  estimation for {MC}-{CDMA} with an antenna array,'' \emph{{IEEE} Trans. Veh.
  Technol.}, vol.~58, no.~8, pp. 4686--4690, Oct. 2009.

\bibitem{fang-ming_han_esprit-like_2005}
F.-M. Han and X.-D. Zhang, ``An {ESPRIT}-like algorithm for coherent {DOA}
  estimation,'' \emph{{IEEE} Antennas Wireless Propag. Lett.}, vol.~4, pp.
  443--446, 2005.

\bibitem{10007797}
F.~Chen, D.~Yang, and S.~Mo, ``A {DOA} estimation algorithm based on
  eigenvalues ranking problem,'' \emph{{IEEE} Trans. Instrum. Meas.}, vol.~72,
  pp. 1--15, 2023.

\bibitem{10041189}
Q.~Guo, Z.~Xin, T.~Zhou, and S.~Xu, ``Off-grid space alternating sparse
  bayesian learning,'' \emph{{IEEE} Trans. Instrum. Meas.}, vol.~72, pp. 1--10,
  2023.

\bibitem{wan_deep_2021}
L.~Wan, Y.~Sun, L.~Sun, Z.~Ning, and J.~J. P.~C. Rodrigues, ``Deep learning
  based autonomous vehicle super resolution {DOA} estimation for safety
  driving,'' \emph{{IEEE} Trans. Intell. Transp. Syst.}, vol.~22, no.~7, pp.
  4301--4315, Jul. 2021.

\bibitem{p_chen_off-grid_2019}
P.~Chen, Z.~Cao, Z.~Chen, and X.~Wang, ``Off-grid {DOA} estimation using sparse
  {Bayesian} learning in {MIMO} radar with unknown mutual coupling,''
  \emph{{IEEE} Trans. Signal Process.}, vol.~67, no.~1, pp. 208--220, Jan.
  2019.

\bibitem{10172104}
S.~Jiang, N.~Fu, Z.~Wei, Z.~Lian, L.~Qiao, and X.~Peng, ``Compressed sampling
  for spectrum measurement and {DOA} estimation with array cooperative {MWC},''
  \emph{{IEEE} Trans. Instrum. Meas.}, vol.~72, pp. 1--14, 2023.

\bibitem{Dai_So_2021}
J.~Dai and H.~C. So, ``Real-valued sparse {Bayesian} learning for {DOA}
  estimation with arbitrary linear arrays,'' \emph{{IEEE} Trans. Signal
  Process.}, vol.~69, pp. 4977--4990, Aug. 2021.

\bibitem{mao_marginal_2021}
Y.~Mao, Q.~Guo, J.~Ding, F.~Liu, and Y.~Yu, ``Marginal likelihood maximization
  based fast array manifold matrix learning for direction of arrival
  estimation,'' \emph{{IEEE} Trans. Signal Process.}, vol.~69, pp. 5512--5522,
  2021.

\bibitem{wan_robust_2022}
J.~Wan, C.~Wang, P.~Shen, H.~Fu, and J.~Zhu, ``Robust and fast super-resolution
  {SAR} tomography of forests based on covariance vector sparse {Bayesian}
  learning,'' \emph{{IEEE} Geosci. Remote Sens. Lett.}, vol.~19, pp. 1--5,
  2022.

\bibitem{wang_alternative_2018}
L.~Wang, L.~Zhao, S.~Rahardja, and G.~Bi, ``Alternative to extended block
  sparse {Bayesian} learning and its relation to pattern-coupled sparse
  {Bayesian} learning,'' \emph{{IEEE} Trans. Signal Process.}, vol.~66, no.~10,
  pp. 2759--2771, May 2018.

\bibitem{5466152}
M.~M. Hyder and K.~Mahata, ``Direction-of-arrival estimation using a mixed
  $\ell _{2,0}$ norm approximation,'' \emph{{IEEE} Trans. Signal Process.},
  vol.~58, no.~9, pp. 4646--4655, 2010.

\bibitem{lu_direction--arrival_2018}
R.~Lu, M.~Zhang, X.~Liu, X.~Chen, and A.~Zhang, ``Direction-of-arrival
  estimation via coarray with model errors,'' \emph{IEEE Access}, vol.~6, pp.
  56\,514--56\,525, 2018.

\bibitem{ruan_parafac_2019}
N.~Ruan, F.~Wen, L.~Ai, and K.~Xie, ``A {PARAFAC} decomposition algorithm for
  {DOA} estimation in colocated {MIMO} radar with imperfect waveforms,''
  \emph{IEEE Access}, vol.~7, pp. 14\,680--14\,688, 2019.

\bibitem{liu_2-d_2021}
S.~Liu, Z.~Zhang, and Y.~Guo, ``\BIBforeignlanguage{en}{2-{D} {DOA} estimation
  with imperfect {L}-shaped array using active calibration},''
  \emph{\BIBforeignlanguage{en}{{IEEE} Commun. Lett.}}, vol.~25, no.~4, pp.
  1178--1182, Apr. 2021.

\bibitem{9173575}
L.~Wan, Y.~Sun, L.~Sun, Z.~Ning, and J.~J. P.~C. Rodrigues, ``Deep learning
  based autonomous vehicle super resolution {DOA} estimation for safety
  driving,'' \emph{{IEEE} Trans. Intell. Transp. Syst.}, vol.~22, no.~7, pp.
  4301--4315, 2021.

\bibitem{8400482}
H.~Huang, J.~Yang, H.~Huang, Y.~Song, and G.~Gui, ``Deep learning for
  super-resolution channel estimation and {DOA} estimation based massive {MIMO}
  system,'' \emph{{IEEE} Trans. Veh. Technol.}, vol.~67, no.~9, pp. 8549--8560,
  2018.

\bibitem{7497454}
H.~Lee, J.~Cho, M.~Kim, and H.~Park, ``{DNN}-based feature enhancement using
  {DOA}-constrained {ICA} for robust speech recognition,'' \emph{{IEEE} Signal
  Process. Lett.}, vol.~23, no.~8, pp. 1091--1095, 2016.

\bibitem{9178434}
T.~N.~T. Nguyen, W.-S. Gan, R.~Ranjan, and D.~L. Jones, ``Robust source
  counting and {DOA} estimation using spatial pseudo-spectrum and convolutional
  neural network,'' \emph{IEEE/ACM Transactions on Audio, Speech, and Language
  Processing}, vol.~28, pp. 2626--2637, 2020.

\bibitem{yuan_unsupervised_2021}
Y.~Yuan, S.~Wu, M.~Wu, and N.~Yuan, ``Unsupervised learning strategy for
  direction-of-arrival estimation network,'' \emph{{IEEE} Signal Process.
  Lett.}, vol.~28, pp. 1450--1454, 2021.

\bibitem{wu_deep_2019}
L.~Wu, Z.~Liu, and Z.~Huang, ``Deep convolution network for direction of
  arrival estimation with sparse prior,'' \emph{{IEEE} Signal Process. Lett.},
  vol.~26, no.~11, pp. 1688--1692, Nov. 2019.

\bibitem{papageorgiou_deep_2021}
G.~Papageorgiou, M.~Sellathurai, and Y.~Eldar, ``Deep networks for
  direction-of-arrival estimation in low {SNR},'' \emph{{IEEE} Trans. Signal
  Process.}, vol.~69, pp. 3714--3729, 2021.

\bibitem{akter_rfdoa-net_2021}
R.~Akter, V.~Doan, T.~Huynh-The, and D.~Kim, ``{RFDOA}-{Net}: {An} efficient
  {ConvNet} for {RF}-based {DOA} estimation in {UAV} surveillance systems,''
  \emph{{IEEE} Trans. Veh. Technol.}, vol.~70, no.~11, pp. 12\,209--12\,214,
  Nov. 2021.

\bibitem{8854868}
L.~Wu, Z.-M. Liu, and Z.-T. Huang, ``Deep convolution network for direction of
  arrival estimation with sparse prior,'' \emph{{IEEE} Signal Process. Lett.},
  vol.~26, no.~11, pp. 1688--1692, 2019.

\bibitem{9328861}
A.~M. Ahmed, U.~S. K. P.~M. Thanthrige, A.~E. Gamal, and A.~Sezgin, ``Deep
  learning for {DOA} estimation in {MIMO} radar systems via emulation of large
  antenna arrays,'' \emph{{IEEE} Commun. Lett.}, vol.~25, no.~5, pp.
  1559--1563, 2021.

\bibitem{9457195}
G.~K. Papageorgiou, M.~Sellathurai, and Y.~C. Eldar, ``Deep networks for
  direction-of-arrival estimation in low {SNR},'' \emph{{IEEE} Trans. Signal
  Process.}, vol.~69, pp. 3714--3729, 2021.

\bibitem{9034077}
A.~M. Elbir, ``{DeepMUSIC: Multiple} signal classification via deep learning,''
  \emph{{IEEE Sensors Lett.}}, vol.~4, no.~4, pp. 1--4, 2020.

\bibitem{8485631}
Z.~Liu, C.~Zhang, and P.~Yu, ``Direction-of-arrival estimation based on deep
  neural networks with robustness to array imperfections,'' \emph{{IEEE} Trans.
  Antennas Propag.}, vol.~66, no.~12, pp. 7315--7327, 2018.

\bibitem{chakrabarty_multi-speaker_2019}
S.~Chakrabarty and E.~A.~P. Habets, ``Multi-speaker {DOA} estimation using deep
  convolutional networks trained with noise signals,'' \emph{IEEE Journal of
  Selected Topics in Signal Processing}, vol.~13, no.~1, pp. 8--21, Mar. 2019.

\bibitem{p_chen_new_2020}
P.~Chen, Z.~Chen, Z.~Cao, and X.~Wang, ``A new atomic norm for {DOA} estimation
  with gain-phase errors,'' \emph{{IEEE} Trans. Signal Process.}, vol.~68, pp.
  4293--4306, 2020.

\bibitem{wang_gridless_2019}
Q.~Wang, X.~Wang, T.~Dou, H.~Chen, and X.~Wu, ``Gridless super-resolution doa
  estimation with unknown mutual coupling,'' in \emph{{ICASSP} 2019 - 2019
  {IEEE} {International} {Conference} on {Acoustics}, {Speech} and {Signal}
  {Processing} ({ICASSP})}.\hskip 1em plus 0.5em minus 0.4em\relax Brighton,
  United Kingdom: IEEE, May 2019, pp. 4210--4214.

\bibitem{govinda_raj_single_2019-1}
A.~Govinda~Raj and J.~H. McClellan, ``Single snapshot super-resolution {DOA}
  estimation for arbitrary array geometries,'' \emph{{IEEE} Signal Process.
  Lett.}, vol.~26, no.~1, pp. 119--123, Jan. 2019.

\bibitem{gong_doa_2022}
Q.~Gong, S.~Ren, S.~Zhong, and W.~Wang, ``{DOA} estimation using sparse array
  with gain-phase error based on a novel atomic norm,'' \emph{Digital Signal
  Processing}, vol. 120, p. 103266, Jan. 2022.

\bibitem{liao_music_2016}
W.~Liao and A.~Fannjiang, ``\BIBforeignlanguage{en}{{MUSIC} for single-snapshot
  spectral estimation: {Stability} and super-resolution},''
  \emph{\BIBforeignlanguage{en}{Applied and Computational Harmonic Analysis}},
  vol.~40, no.~1, pp. 33--67, Jan. 2016.

\bibitem{govinda_raj_single_2019}
A.~Govinda~Raj and J.~H. McClellan, ``Single snapshot super-resolution {DOA}
  estimation for arbitrary array geometries,'' \emph{{IEEE} Signal Process.
  Lett.}, vol.~26, no.~1, pp. 119--123, Jan. 2019.

\bibitem{wei_gridless_2020}
Z.~Wei, W.~Wang, F.~Dong, and Q.~Liu, ``Gridless one-bit direction-of-arrival
  estimation via atomic norm denoising,'' \emph{{IEEE} Commun. Lett.}, vol.~24,
  no.~10, pp. 2177--2181, Oct. 2020.

\bibitem{zai_yang_enhancing_2016}
Z.~Yang and L.~Xie, ``\BIBforeignlanguage{en}{Enhancing sparsity and resolution
  via reweighted atomic norm minimization},''
  \emph{\BIBforeignlanguage{en}{{IEEE} Trans. Signal Process.}}, vol.~64,
  no.~4, pp. 995--1006, Feb. 2016.

\bibitem{z_yang_orthonormal_2011}
{Z. Yang}, {C. Zhang}, {J. Deng}, and {W. Lu}, ``Orthonormal expansion
  $\ell_{1}$-minimization algorithms for compressed sensing,'' \emph{{IEEE}
  Trans. Signal Process.}, vol.~59, no.~12, pp. 6285--6290, Dec. 2011.

\bibitem{z_tan_joint_2014}
{Z. Tan}, {P. Yang}, and {A. Nehorai}, ``Joint {Sparse} {Recovery} {Method} for
  {Compressed} {Sensing} {With} {Structured} {Dictionary} {Mismatches},''
  \emph{{IEEE} Trans. Signal Process.}, vol.~62, no.~19, pp. 4997--5008, Oct.
  2014.

\bibitem{g_yu_statistical_2011}
{G. Yu} and {G. Sapiro}, ``Statistical compressed sensing of {Gaussian} mixture
  models,'' \emph{{IEEE} Trans. Signal Process.}, vol.~59, no.~12, pp.
  5842--5858, Dec. 2011.

\bibitem{aghababaiyan_high-precision_2020}
K.~Aghababaiyan, V.~Shah-Mansouri, and B.~Maham, ``High-precision {OMP}-based
  direction of arrival estimation scheme for hybrid non-uniform array,''
  \emph{{IEEE} Commun. Lett.}, vol.~24, no.~2, pp. 354--357, Feb. 2020.

\bibitem{lin_single_2021}
M.~Lin, M.~Xu, X.~Wan, H.~Liu, Z.~Wu, J.~Liu, B.~Deng, D.~Guan, and S.~Zha,
  ``Single sensor to estimate {DOA} with programmable metasurface,''
  \emph{{IEEE Internet of Things Journal}}, vol.~8, no.~12, pp.
  10\,187--10\,197, Jun. 2021.

\bibitem{chen_source_2019}
Y.~Chen, W.~Wang, Z.~Wang, and B.~Xia, ``A source counting method using
  acoustic vector sensor based on sparse modeling of {DOA} histogram,''
  \emph{{IEEE} Signal Process. Lett.}, vol.~26, no.~1, pp. 69--73, Jan. 2019.

\bibitem{izacard2019data}
G.~Izacard, S.~Mohan, and C.~Fernandez-Granda, ``Data-driven estimation of
  sinusoid frequencies,'' in \emph{NeurIPS}, 2019.

\end{thebibliography}
 
\begin{IEEEbiography}[{\includegraphics[width=1in,height=1.25in,clip,keepaspectratio]{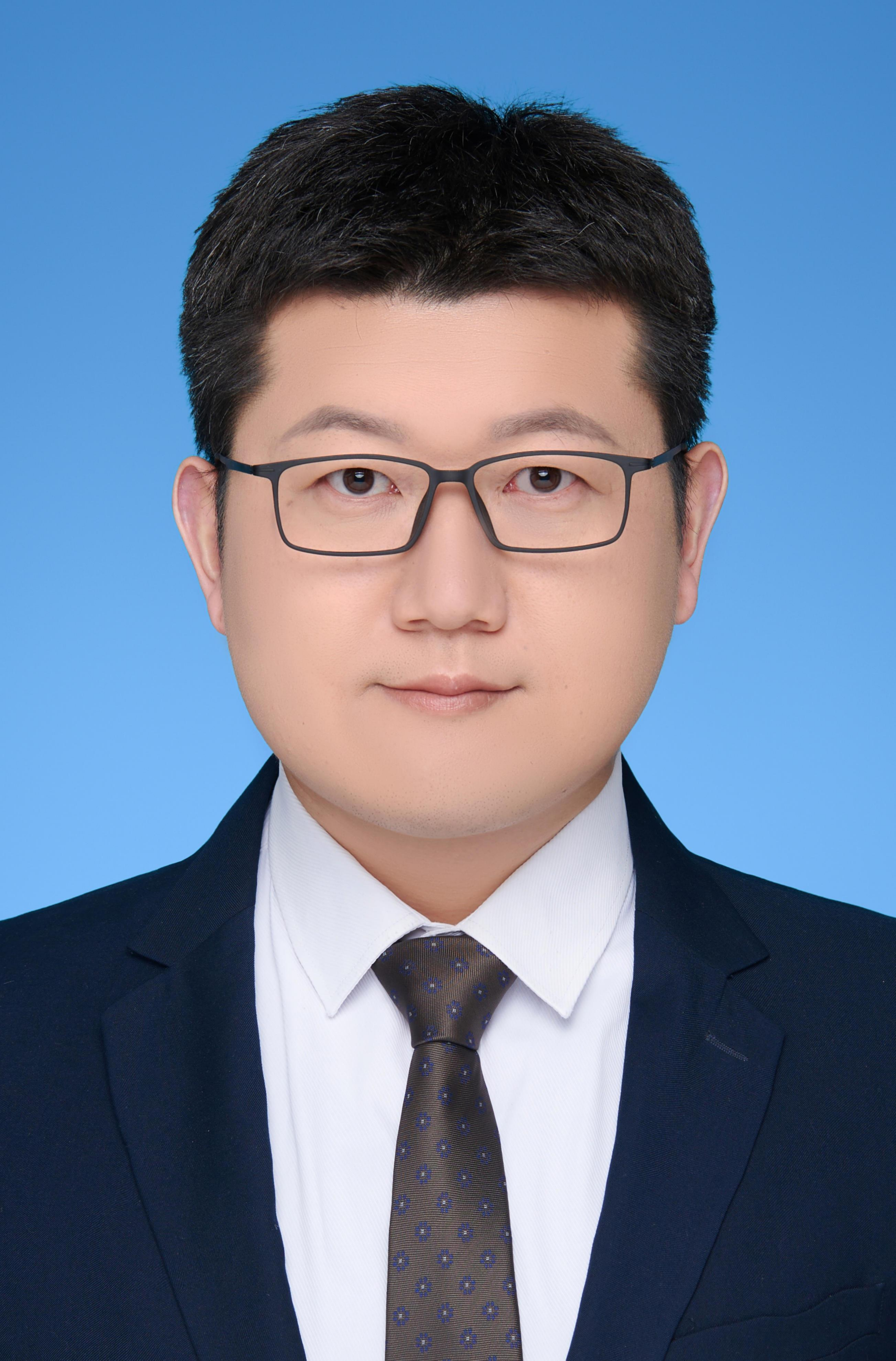}}]{Peng Chen (Seinor Member, IEEE)} received the B.E. and Ph.D. degrees from the School of Information Science and Engineering, Southeast University, Nanjing, China, in 2011 and 2017 respectively. From March 2015 to April 2016, he was a Visiting Scholar with the Department of Electrical Engineering, Columbia University, New York, NY, USA. He is currently an Associate Professor with the State Key Laboratory of Millimeter Waves, Southeast University. His research interests include target localization, super-resolution reconstruction, and array signal processing. He is a Jiangsu Province Outstanding Young Scientist. He has served as an IEEE ICCC Session Chair, and won the Best Presentation Award in 2022 (IEEE ICCC). He was invited as a keynote speaker at the IEEE ICET in 2022. He was recognized as an exemplary reviewer for IEEE WCL in 2021, and won the Best Paper Award at IEEE ICCCCEE in 2017.
\end{IEEEbiography}
 
\begin{IEEEbiography}[{\includegraphics[width=1in,height=1.25in,clip,keepaspectratio]{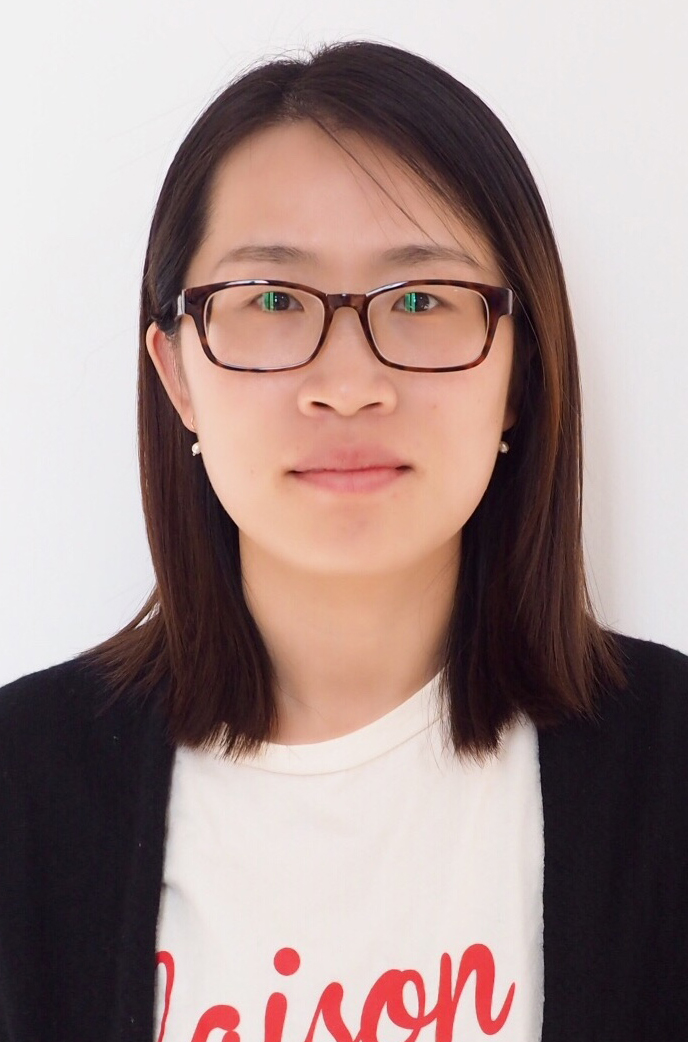}}]{Zhimin Chen (Member, IEEE)} received the Ph.D. degree in information and communication engineering from the School of Information Science and Engineering, Southeast University, Nanjing, China in 2015. Since 2015, she is currently an associate professor at Shanghai Dianji University, Shanghai, China. From 2021,  she is also a Visiting Scholar in  the Department of Electronic and Information Engineering, The Hong Kong Polytechnic University, Hong Kong.  Her research interests include array signal processing, vehicle communications and millimeter-wave communications. 
\end{IEEEbiography}

\begin{IEEEbiography}[{\includegraphics[width=1in,height=1.25in,clip,keepaspectratio]{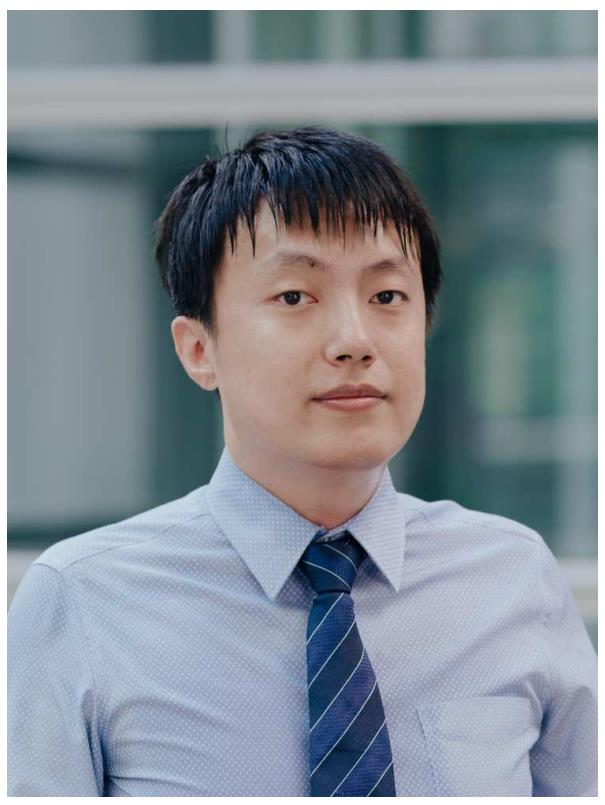}}]{Liang Liu (S'14-M'15)}  received the B.Eng. degree from the Tianjin University, China, in 2010, and the Ph.D. degree from the National University of Singapore in 2014. He is currently an Assistant Professor in the Department of Electronic and Information Engineering at the Hong Kong Polytechnic University. Before that, he was a Research Fellow in the Department of Electrical and Computer Engineering at National University of Singapore from 2017 to 2018, and a Postdoctoral Fellow in the Department of Electrical and Computer Engineering at the University of Toronto from 2015 to 2017. His research interests include the next generation cellular technologies and machine-type communications for the Internet of Things. He was the recipient of the 2021 IEEE Signal Processing Society Best Paper Award, the 2017 IEEE Signal Processing Society Young Author Best Paper Award, and the Best Paper Award from the 2011 International Conference on Wireless Communications and Signal Processing. He was recognized by Clarivate Analytics as a Highly Cited Researcher in 2018. He is an editor for IEEE Transactions on Wireless Communications, and was a leading guest editor for IEEE Wireless Communications special issue on “Massive Machine-Type Communications for IoT”.
\end{IEEEbiography} 

\begin{IEEEbiography}[{\includegraphics[width=1in,height=1.25in,clip,keepaspectratio]{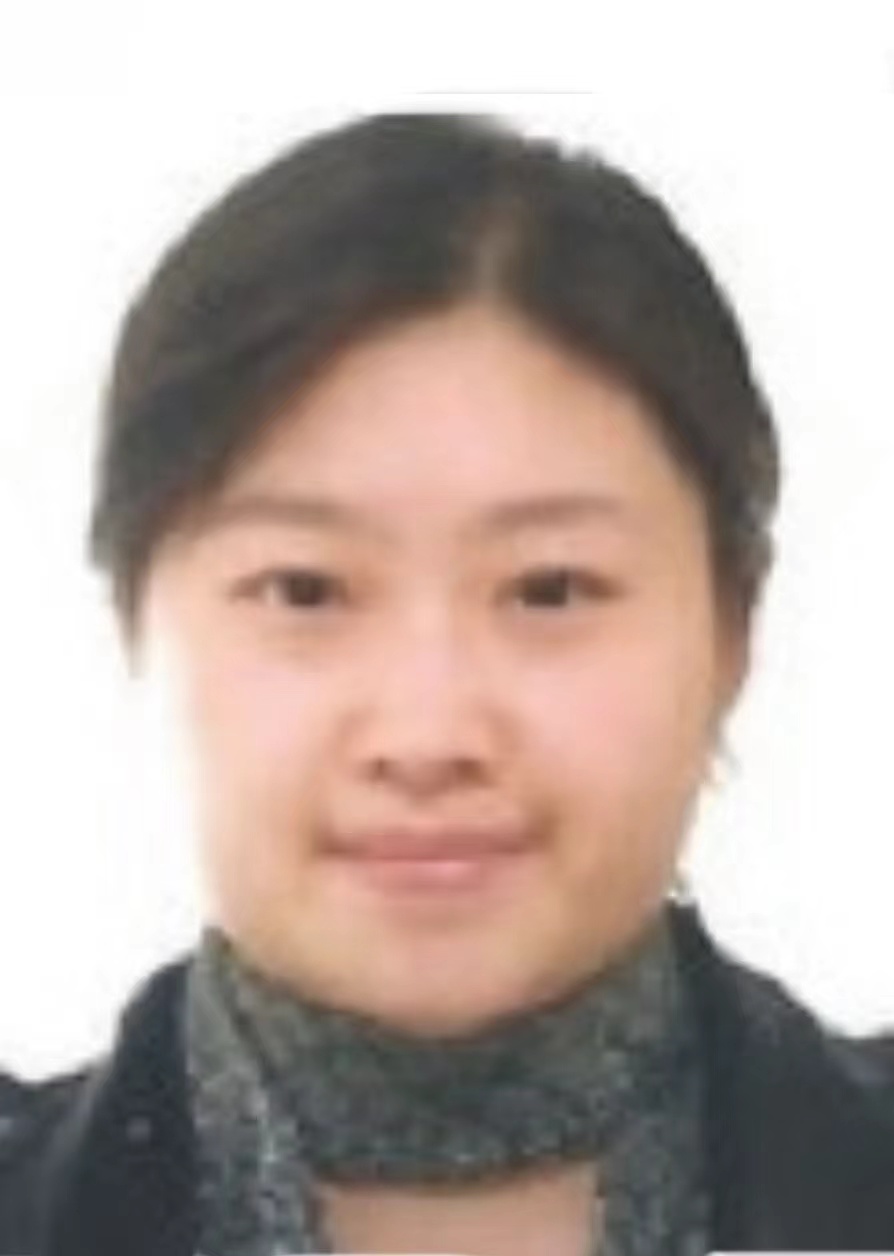}}]{Yun Chen (Member, IEEE)} received the B.Sc. degree from UESTC in 2000 and the Ph.D. degree from Fudan University in 2007. 

In 2007, she joined Fudan University, where she has been with the faculty since March 2008. She has been with Fudan University as an Associate Professor of the State Key Laboratory of ASIC and System. She has published more than 60 articles in such international journals and conferences as IEEE ASSCC, IEEE Transactions ON CAS, IEEE Transactions ON Communication, IEEE ASP-DAC, IEEE ICASSP, ICC, and ISCAS. She applied for more than 20 patents. Her research fields include base-band processing technologies for wireless communication and ultra-low power FEC IC design. She also serves as a Chair Secretary of the Shanghai Chapter of IEEE SSCS, a TPC Member of ASSCC, a Co-Chair of the Circuit System Division, Chinese Institute of Electronics, and a member of the Steering Committee of SIPS and the ASICON technical committee.
\end{IEEEbiography} 
	
\begin{IEEEbiography}
	[{\includegraphics[width=1in,height=1.25in,clip,keepaspectratio]{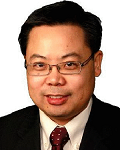}}]{Xianbin Wang (Fellow, IEEE)} received the Ph.D. degree in electrical and computer engineering from National University of Singapore, in 2001. He is a Professor and Tier-I Canada Research Chair at Western University, Canada. Prior to joining Western University, he was with Communications Research Centre (CRC) Canada as a Research Scientist/Senior Research Scientist between July 2002 and December 2007. From January 2001 to July 2002, he was a System Designer with STMicroelectronics, where he was responsible for the system design of DSL and Gigabit Ethernet chipsets. His current research interests include 5G technologies, Internet of Things, communications security, machine learning and locationing technologies. Dr. Wang has over 300 peer-reviewed journal and conference papers, in addition to 26 granted and pending patents and several standard contributions. Dr. Wang is a Fellow of Canadian Academy of Engineering and an IEEE Distinguished Lecturer. He has received many awards and recognitions, including Canada Research Chair, CRC Presidents Excellence Award, Canadian Federal Government Public Service Award, Ontario Early Researcher Award and five IEEE Best Paper Awards. He currently serves as an Editor/Associate Editor for IEEE Transactions on Communications, IEEE Transactions on Broadcasting, and IEEE Transactions on Vehicular Technology and he was also an Associate Editor for IEEE Transactions on Wireless Communications between 2007 and 2011, and IEEE Wireless Communications Letters between 2011 and 2016. Dr. Wang was involved in many IEEE conferences including GLOBECOM, ICC, VTC, PIMRC, WCNC, and CWIT, in different roles such as symposium chair, tutorial instructor, track chair, session chair and TPC co-chair.
\end{IEEEbiography}
\end{document}